\documentclass[a4paper,11pt]{article}
\pdfoutput=1 

\usepackage{jheppub}
\usepackage[T1]{fontenc}

\newcommand\T{\rule{0pt}{2.6ex}}       
\newcommand\TT{\rule{0pt}{3.0ex}}       
\newcommand\B{\rule[-1.2ex]{0pt}{0pt}} 
\newcommand{\nn}{\nonumber \\} 
\newcommand{\bea}{\begin{eqnarray}}  
\newcommand{\eea}{\end{eqnarray}}  
\newcommand{\Raw}{\Rightarrow}
\newcommand{\ba}{\begin{array}}
\newcommand{\ea}{\end{array}}
\newcommand{\dis}{\displaystyle}
\newcommand{\lag}{{\cal L}}

\title{
Lepton Flavor Changing Higgs decays in\\ the Littlest Higgs Model with T-parity
}

\author[a]{Francisco del Aguila,}
\author[b]{Lluis Ametller,}
\author[a]{Jose Ignacio Illana,}
\author[a]{Jose Santiago,}
\author[c]{Pere Talavera,}
\author[a]{and Roberto Vega-Morales~}

\affiliation[a]{CAFPE and Departamento de F{\'\i}sica Te{\'o}rica y del Cosmos, Universidad
de Granada, E\textendash{}18071 Granada, Spain}
\affiliation[b]{Departament de F\'{\i}sica, Universitat Polit{\`e}cnica de Catalunya, 
E-08034 Barcelona, Spain}
\affiliation[c]{Institut de Ciencies del Cosmos, Universitat de
  Barcelona (IEEC-UB), Marti i Franques 1, Barcelona 08028, Spain}   

\emailAdd{faguila@ugr.es}
\emailAdd{lluis.ametller@upc.edu}
\emailAdd{jillana@ugr.es}
\emailAdd{jsantiago@ugr.es}
\emailAdd{ptalavera@iese.edu}
\emailAdd{rvegamorales@ugr.es}

\abstract{We calculate loop induced lepton flavor violating Higgs decays in the Littlest Higgs model with T-parity. 
We find that a finite amplitude is obtained only when all contributions from the T-odd lepton sector are included. 
This is in contrast to lepton flavor violating processes mediated by
gauge bosons where the partners of the right-handed mirror leptons can
be decoupled from the spectrum.   
These partners are necessary to cancel the divergence in the Higgs
mass introduced by the mirror leptons but are otherwise unnecessary
and assumed to be decoupled in previous phenomenological studies. 
Furthermore, as we emphasize, including the partner leptons in the
spectrum also introduces a new source of lepton flavor violation via
their couplings to the physical pseudo-Goldstone electroweak triplet
scalar. 
Although this extra source also affects lepton flavor changing gauge
transitions, it decouples from these amplitudes in the limit of heavy
mass for the partner leptons. 
We find that the corresponding Higgs branching ratio into taus and
muons can be as large as $\sim 0.2 \times 10^{-6}$ for T-odd masses of
the order a few TeV, a demanding challenge even for the high
luminosity LHC.}

\begin{document}
\maketitle
\flushbottom

\section{Introduction}
\label{Introduction}

The discovery at the LHC
\cite{Chatrchyan:2012xdj,Aad:2012tfa} 
of a Standard Model (SM) like Higgs 
\cite{Falkowski:2013dza,Khachatryan:2014kca,Khachatryan:2016vau} 
appears to have settled the nature of the electroweak symmetry breaking 
(EWSB) mechanism, but the lack of any signs of physics beyond the SM 
still leaves unsolved the apparent fine tuning problem of the electroweak scale. 
While lack of observed new physics strongly constrains the most natural models 
addressing the hierarchy problem, room still exists for a solution at $\sim$ TeV 
which is perhaps observable at the LHC.

Although supersymmetry is the most well known solution, Little Higgs models 
\cite{ArkaniHamed:2001nc} also offer an elegant solution and will be further 
tested in the coming years at the LHC. 
(For a review see Refs. \cite{Schmaltz:2005ky,Han:2005ru,Perelstein:2005ka}.) 
A popular, phenomenologically viable 
realization of this class of models is the Littlest Higgs model with T-parity 
\cite{Cheng:2003ju,Cheng:2004yc,Low:2004xc} 
(LHT), which is protected at tree level against constraints from electroweak 
precision \cite{Hubisz:2004ft,Hubisz:2005tx} data (EWPD). 
Restricting ourselves to `non-linear' 
Callan-Coleman-Wess-Zumino \cite{Callan:1969sn} 
constructions \cite{Low:2004xc,Hill:2007zv}, the LHT 
possesses a global $SU(5)$ symmetry broken spontaneously at a scale 
$f \sim$ TeV down to an $SO(5)$ subgroup. This gives rise to a Goldstone sector 
containing the SM Higgs doublet as well as an electroweak triplet with 
zero vacuum expectation value (\emph{vev}).\footnote{
This is true only if T-parity remains exact, but a tiny \emph{vev} along the neutral direction of this pseudo-Goldstone scalar 
triplet can be invoked to give very small Majorana masses to the SM neutrinos 
\cite{Han:2005nk}.}
Two different $SU(2)\times U(1)$ subgroups of $SU(5)$ are gauged with equal 
strength and broken spontaneously down to a diagonal subgroup which is 
identified with the SM gauge group, while the broken gauge symmetries 
lead to a set of massive vector bosons with mass $\sim f$. 
Utilizing a $Z_2$ automorphism inherited from the global symmetry breaking 
structure, a `T-parity' can be defined for the gauge sector 
under which all SM particles are even 
and with mass scale $v \sim 246$~GeV and (almost) all new particles are odd 
with mass scale $f$.

There have been many phenomenological studies of this type of Little Higgs models 
\cite{Hubisz:2005bd,Chen:2006cs,Blanke:2006sb,Buras:2006wk,Belyaev:2006jh,Hill:2007nz,Han:2008wb,Goto:2008fj,delAguila:2008zu,Blanke:2009am,delAguila:2010nv,Goto:2010sn,Zhou:2012cja}. 
A number of these have examined specifically the possibility of lepton flavor violation (LFV) 
in processes at LEP and LHC or other experiments
~\cite{Hubisz:2005bd,delAguila:2008zu,Blanke:2009am,delAguila:2010nv,Goto:2010sn}.\footnote{
Similar studies for the Simplest Little
Higgs model can be found in Refs.
\cite{Kaplan:2003uc,Schmaltz:2004de,delAguila:2011wk,Lami:2016mjf}, and references therein. 
} 
The underlying source of this LFV comes from the Yukawa sectors responsible 
for generating the masses of the light SM leptons and the heavy T-odd `mirror' leptons.
In particular, it arises from the rotations in flavor space which are necessary to 
diagonalize their mass matrices.
However, as we emphasize in this work, this is not the only possible source of LFV in the LHT. 
Apart from the T-odd mirror leptons, which acquire a mass $\sim f$ after the $SU(5)$ 
breaking, there are additional T-odd lepton doublets (also vector-like) required to maintain the $SO(5)$ 
global symmetry protecting the Higgs mass from dangerous divergences \cite{Cheng:2004yc}. 
These do not acquire a mass from the $SU(5)$ breaking and are often not necessary 
for understanding the essential aspects of the LHT or much of its phenomenology. 
Thus, in virtually all phenomenological studies of LFV in the LHT, it is assumed that 
these additional leptons completing the $SO(5)$ right-handed representations are decoupled 
from the spectrum, though light enough to not reintroduce a fine tuning in the Higgs mass 
for which they serve as a cutoff at two loops \cite{Cheng:2004yc}. 
However, as we examine more closely below, such a decoupling does not hold when dealing with Higgs decays. 
Moreover, since they do not acquire a mass from the same 
mechanism which generates masses for the mirror leptons, these partner leptons must be given separate masses. 
These masses, which softly break the global $SO(5)$, will themselves carry generation indices and introduce additional 
flavor structure in the Yukawa sector.

The rotations of these additional lepton fields, required to diagonalize their mass matrix, are independent of the previous rotations needed to diagonalize the SM and mirror lepton masses and introduce a potential new source of LFV. 
In this sense, previous studies which assume these partner leptons are heavy are not exhaustive. 
Furthermore, since the new contributions from the heavy partner leptons are necessary to render LFV Higgs decay amplitudes finite, any previous estimates of these decays which neglect them must be recomputed to give an unambiguous prediction. 
In contrast, LFV effects mediated by photons or $Z$ bosons which explicitly depend on these new rotations become negligible in the limit of heavy partner leptons and in particular, there is no divergence introduced when the partner leptons are neglected.
So while LFV effects, in $h\to\tau\mu$ for instance, from the partner leptons can be sizable even when they are heavy compared to the $SU(5)$ breaking scale $f$, they do decouple in processes mediated by gauge bosons such as $Z\to\tau\mu$. 
This implies that $h \to\tau\mu$ decays, for which constraints from LHC are weaker 
\cite{Khachatryan:2015kon}, could be significantly larger than naively expected given constraints on LFV from processes mediated by photons or $Z$ bosons.

We compute the one-loop contributions to $h \to\tau\mu$ decays and show that the T-odd particles sum to a finite amplitude, but only once all T-odd leptons are included. 
We emphasize this here because the contributions from the mirror leptons, which are not decoupled in studies of $\gamma$ and $Z$ mediated LFV transitions, are not finite by themselves in $h\to\tau\mu$.\footnote{
Against what is stated in Refs. \cite{Yang:2016hrh,Chamorro-Solano:2016ugt}.}
This is in contrast to $Z\to\tau\mu$ for example where they are indeed finite. 
Furthermore, as we examine below, including the partner leptons in the spectrum necessarily introduces new couplings to the physical pseudo-Goldstone electroweak triplet scalar which will enter into LFV amplitudes and which depend explicitly on the rotations of the heavy partner lepton fields needed to diagonalize their mass matrices.
Although this extra source also enters in gauge boson mediated LFV amplitudes, its effects decouple in the limit of heavy mass for the partner leptons.
For $h\to\tau\mu$ their contribution, through which the new LFV source enters, does not decouple in the limit of heavy partner lepton masses 
and instead grows logarithmically.
These logarithmic corrections are indicative of a soft breaking of the global $SO(5)$ by the partner lepton masses and are related to corrections to the Higgs mass.
Therefore, the partner lepton masses cannot be taken arbitrarily large without reintroducing a fine tuning problem in the Higgs mass at two loops.

To be self contained and to fix our notation we review in the next section the LHT and provide the new set of Feynman rules necessary for computing LFV Higgs decays, including the couplings involving the heavy partner leptons.
The interested reader can go directly to Section 3 where we present
the calculation of the $h \to \overline{\ell} \ell^\prime$ amplitude.   
For completeness, the detailed expression of the contribution 
independent of the new LFV source in the $h\to \overline{\ell}
\ell^\prime$ amplitude  
is gathered in the Appendix. 
Section 4 is devoted to a short phenomenological discussion, including
the effect of the proper redefinition of the final mass
eigenstates. We also point out the main parametric limits for the
$h \rightarrow \tau \mu$ branching ratio, which
can be raised up to $\sim 0.2 \times 10^{-6}$ in the limit of large mirror and partner lepton masses, 
becoming a demanding challenge even for the high luminosity LHC. 
A detailed phenomenological discussion will be presented elsewhere \cite{inpreparation}, 
together with a reanalysis of LFV processes mediated by neutral gauge bosons which includes the new contributions from the right handed partner leptons. 
The final section is devoted to our conclusions. 

\section{Reviewing the LHT}
\label{Model}

Here we summarize the LHT to fix our notation and provide the new Feynman rules necessary for the one loop calculation of the $h\to\tau\mu$ amplitude. 
Since the amplitude is higher order in an operator expansion in SM fields it is in general suppressed by $v^2/f^2$ and clearly decouples when $f$ goes to infinity. 
Thus, in general the Feynman rules involving the heavy T-odd particles must also be worked out to this order.
Though we attempt to be as self-contained as possible, we closely follow the presentation in 
Refs. \cite{Blanke:2006eb,delAguila:2008zu}. 

The LHT is a non-linear $\sigma$ model with a $SU(5)$ global symmetry broken down to $SO(5)$ by the
 {\emph{vev}} of a $5 \times 5$ symmetric tensor,  
\bea
\Sigma_0=\left(\ba{ccc} {\bf 0}_{2\times2} & 0 & {\bf 1}_{2\times2} \\
                         0 & 1 &0 \\ 
                         {\bf 1}_{2\times2} & 0 & {\bf 0}_{2\times 2}\ea\right).
\label{Sigma}
\eea
The 10 unbroken $SU(5)$ generators $T^a$, which leave invariant $\Sigma_0$ 
and hence satisfy the equality $T^a \Sigma_0 + \Sigma_0 (T^a)^T = 0$, 
generate the $SO(5)$ algebra; 
whereas the 14 broken $SU(5)$ generators $X^a$, 
which fulfill the relation $X^a \Sigma_0 - \Sigma_0 (X^a)^T = 0$,  
expand the Goldstone matrix $\Pi=\pi^a X^a$ parameterizing the 
$5 \times 5$ symmetric tensor 
\bea
\Sigma(x)={\rm e}^{i\Pi/f}\Sigma_0{\rm e}^{i\Pi^T/f}={\rm e}^{2i\Pi/f}\Sigma_0\ , 
\label{Sigmaparameterization}
\eea
with $f$ the scale of new physics (NP). 
It is important to note that this breaking fixes the embedding of $SO(5)$ in $SU(5)$, 
with the fundamental representation of $SU(5)$ reducing to the 
defining (real) representation of $SO(5)$, both of dimension five.

An $SU(2)_1 \times SU(2)_2 \times U(1)_1 \times U(1)_2$ subgroup of the $SU(5)$ is gauged and generated by
\bea
Q_1^a=\frac{1}{2}\left(\ba{ccc}\sigma^a & 0 & 0 \\ 0 & 0 & 0 \\ 0 & 0 &  {\bf 0}_{2\times2} \ea\right),\quad
Y_1=\frac{1}{10}{\rm diag}(3,3,-2,-2,-2)\ , \nonumber 
\\ 
\\
Q_2^a=\frac{1}{2}\left(\ba{ccc}{\bf 0}_{2\times2} & 0 & 0 \\ 0 & 0 & 0 \\ 0 & 0 & -\sigma^{a*} \ea\right),\quad
Y_2=\frac{1}{10}{\rm diag}(2,2,2,-3,-3)\ , \nonumber
\eea
with $\sigma^a$ the three Pauli matrices. The \emph{vev} in Eq. (\ref{Sigma}) 
breaks this gauge group down to the SM gauge group $SU(2)_L \times U(1)_Y$, generated 
by the combinations $\{Q_1^a+Q_2^a,\ Y_1+Y_2\}\subset\{T^a\}$. The orthogonal 
combinations are a subset of the broken generators, 
$\{Q_1^a-Q_2^a,\ Y_1-Y_2\}\subset\{X^a\}$.
Thus, the Goldstone matrix
\bea
\Pi=\left(\ba{ccccc}
-\dis\frac{\omega^0}{2}-\frac{\eta}{\sqrt{20}} & -\dis\frac{\omega^+}{\sqrt{2}} & -i\dis\frac{\pi^+}{\sqrt{2}} & -i\Phi^{++} & -i\dis\frac{\Phi^+}{\sqrt{2}} \\
-\dis\frac{\omega^-}{\sqrt{2}} & \dis\frac{\omega^0}{2}-\frac{\eta}{\sqrt{20}} & \dis\frac{v+h+i\pi^0}{2} & -i\dis\frac{\Phi^+}{\sqrt{2}} & \dis\frac{-i\Phi^0+\Phi^P}{\sqrt{2}} \\
i\dis\frac{\pi^-}{\sqrt{2}} & \dis\frac{v+h-i\pi^0}{2} & \sqrt{\dis\frac{4}{5}}\eta & -i\dis\frac{\pi^+}{\sqrt{2}} &  \dis\frac{v+h+i\pi^0}{2} \\
i\Phi^{--} & i\dis\frac{\Phi^-}{\sqrt{2}} & i\dis\frac{\pi^-}{\sqrt{2}} & -\dis\frac{\omega^0}{2}-\frac{\eta}{\sqrt{20}} & -\dis\frac{\omega^-}{\sqrt{2}} \\
i\dis\frac{\Phi^-}{\sqrt{2}} & \dis\frac{i\Phi^0+\Phi^P}{\sqrt{2}} &  \dis\frac{v+h-i\pi^0}{2} & -\dis\frac{\omega^+}{\sqrt{2}} & \dis\frac{\omega^0}{2}-\frac{\eta}{\sqrt{20}}
\ea\right)
\label{goldstones}
\eea
decomposes into the SM Higgs doublet $\phi^T=(-i\pi^+,
(v+h+i\pi^0)/\sqrt{2})^T$, 
a complex SU(2)$_L$ triplet $\Phi$, and the 
longitudinal modes of the heavy gauge fields $\omega^\pm, \omega^0$ and $\eta$.\footnote{
In the following we use for the SM fields and couplings 
the conventions in Ref.~\cite{Denner:1991kt}. In particular,
$\phi^+=-i\pi^+$, $\phi^0=\pi^0$.}

\vskip 0.2cm
\noindent
{\it Gauge Lagrangian}. 
T-parity is introduced to make the new heavy particles T-odd, keeping  
the SM fields T-even (invariant).
Its action on the gauge fields $G_i$ exchanges the two gauge groups 
$SU(2)_1 \times U(1)_1$ and $SU(2)_2 \times U(1)_2$, 
\bea
G_1\stackrel{{\rm T}}{\longleftrightarrow} G_2\ .
\eea
T-parity then requires the gauge couplings to be equal leading to the gauge Lagrangian 
\bea
\lag_G &=& \sum_{j=1}^2\left[-\frac{1}{2}{\rm Tr}
\left(\widetilde W_{j\mu\nu}\widetilde W_j^{\mu\nu}\right)
               -\frac{1}{4}B_{j\mu\nu}B_j^{\mu\nu}\right], 
\eea
which is T parity and gauge invariant where we have defined
\bea
\widetilde W_{j\mu} = W_{j\mu}^a Q_j^a\ , \quad
\widetilde W_{j\mu\nu}=
\partial_\mu\widetilde W_{j\nu}-\partial_\nu\widetilde W_{j\mu}
-i g\left[\widetilde W_{j\mu},\widetilde W_{j\nu}\right],\quad
B_{j\mu\nu}=\partial_\mu B_{j\nu}-\partial_\nu B_{j\mu} 
\nonumber\\
\eea
and repeated indices are understood to be summed.
The SM gauge bosons are the T-even combinations multiplying 
the unbroken gauge generators,  
\bea
W^\pm=\frac{1}{2}[(W_1^1+W_2^1)\mp i (W_1^2+W_2^2)]\ ,\quad
W^3=\frac{W_1^3+W_2^3}{\sqrt{2}}\ ,\quad
B=\frac{B_1+B_2}{\sqrt{2}}\ ; 
\label{GaugeTeven}
\eea
whereas the heavy gauge bosons are the T-odd combinations 
\bea
W_H^\pm=\frac{1}{2}[(W_1^1-W_2^1)\mp i (W_1^2-W_2^2)]\ ,\quad
W_H^3=\frac{W_1^3-W_2^3}{\sqrt{2}}\ ,\quad
B_H=\frac{B_1-B_2}{\sqrt{2}}\ . 
\label{GaugeTodd}
\eea

\vskip 0.2cm
\noindent 
{\it Scalar Lagrangian}. 
Likewise, to keep the SM Higgs doublet T-even and 
make the remaining Goldstone fields T-odd, we define the 
T action on the scalar fields  
\bea
\Pi\stackrel{{\rm T}}{\longrightarrow}-\Omega\Pi\Omega\ ,\quad
\Omega={\rm diag}(-1,-1,1,-1,-1)\ ,  
\label{Pi}
\eea
where $\Omega$ is an element of the center of the gauge group
commuting with $\Sigma_0$ but not with the full global symmetry.\footnote{
Note that we have reversed the sign of $\Omega$ as compared to the literature, 
to make it a group element.}
Hence, 
\bea
\Sigma\stackrel{{\rm T}}{\longrightarrow}
\widetilde\Sigma=\Omega\Sigma_0\Sigma^\dagger\Sigma_0\Omega \ ,
\label{SigmaT}
\eea
while the scalar kinetic Lagrangian is given by
\bea
\lag_{S} &=& \frac{f^2}{8}{\rm Tr}\left[(D_\mu\Sigma)^\dagger(D^\mu\Sigma)\right],
\label{LS}
\eea
with the covariant derivative defined as
\bea
\label{derivative}
D_\mu\Sigma=\partial_\mu\Sigma-\sqrt{2}i\sum_{j=1}^2
\left[gW_{j\mu}^a(Q_j^a\Sigma+\Sigma Q_j^{aT})-g'B_{j\mu}(Y_j\Sigma+\Sigma Y_j^T)\right].
\eea

\vskip 0.2cm
\noindent 
{\it Fermionic Lagrangian}. 
Implementing T-parity in the fermionic sector of the Littlest Higgs model is less straightforward. 
Three types of couplings are needed to give masses to all leptons in the model. 
It is the misalignment between these couplings which results in the two sources of LFV 
which are beyond the SM.\footnote{
LFV is highly suppressed in the SM due to the tiny neutrino mass which we take them as massless throughout this work.}

Following Refs.~\cite{Cheng:2004yc,Low:2004xc} for each SM left-handed ($L$) lepton doublet 
we introduce two incomplete $SU(5)$ multiplets in fundamental representations: 
\bea
\Psi_1=\left(\ba{c} - i \sigma^2 l_{1L} \\ 0 \\ 0 \ea\right),\quad
\Psi_2=\left(\ba{c} 0 \\ 0 \\ - i \sigma^2 l_{2L} \ea\right) , 
\label{Psimultiplets}
\eea
where $l_{rL}=\left(\ba{c} \nu_{rL} \\ \ell_{rL}\ea\right),\ r=1,2$; and 
\bea
\Psi_1\longrightarrow V^*\Psi_1\ ,\quad
\Psi_2\longrightarrow V\Psi_2\ ,
\eea
under the $SU(5)$ transformation $V$. 
T-parity is then defined 
\bea
\Psi_1\stackrel{{\rm T}}{\longleftrightarrow}\Omega\Sigma_0\Psi_2\ ,
\label{PsiT}
\eea
where the T-even combination is given by $\Psi_1+\Omega\Sigma_0\Psi_2$ and identified with the SM left-handed lepton doublet, up to the proper normalization. 
The orthogonal combination $\Psi_1-\Omega\Sigma_0\Psi_2$ defines a second left-handed lepton doublet which is T-odd (see Eqs. (\ref{Psimultiplets}) and (\ref{PsiT})), and which must be paired to a right-handed ($R$) `mirror' lepton doublet $l_{H R}$ in order to obtain a large (vector-like) mass of ${\cal O} (f)$. ($H$ stands for heavy.)

The mirror leptons introduce divergences into the Higgs mass as well as the $h\to\tau\mu$ amplitude which we examine below.
These divergences must be cancelled by introducing additional right handed `partner' leptons $\tilde\psi_R$, along with $l_{H R}$ and an additional singlet $\chi_R$,  to form a complete $SO(5)$ multiplet which we define as:  
\bea 
\Psi_R =\left(\ba{c} \tilde\psi_R \\ \chi_R \\ - i \sigma^2 l_{H R} \ea\right) , \quad 
\Psi_R&\longrightarrow U\Psi_R\ . 
\label{complete}
\eea
$\Psi_R$ transforms under T-parity 
\bea
\Psi_R\stackrel{{\rm T}}{\longrightarrow}\Omega\Psi_R \ . 
\label{PsiRT}
\eea
The mirror leptons obtain their masses through the non-linear Yukawa Lagrangian: 
\bea
\lag_{Y_H} = -\kappa f \left(\overline\Psi_2\xi+ 
\overline\Psi_1\Sigma_0\xi^\dagger\right)\Psi_R
+{\rm h.c.}\ ,
\label{mirror}
\eea
where $\xi={\rm e}^{i\Pi/f}$. 
This is T-invariant, since Eq. (\ref{Pi}) implies 
\bea
\xi \stackrel{{\rm T}}{\longrightarrow} \Omega\xi^\dagger\Omega \ ,    
\eea
as well as invariant under global $SU(5)$ transformations,
\bea
\Sigma=\xi^2\Sigma_0\longrightarrow V\Sigma V^T \quad 
\Raw \quad \xi\longrightarrow V\xi U^\dagger \equiv U\xi\Sigma_0 V^T\Sigma_0 \ ,
\eea
where $V$ is the global $SU(5)$ transformation while $U$ takes values in the $SO(5)$ Lie algebra and is a function of $V$ and $\Pi$. 
We note that the gauge singlet $\chi_R$ is T-even 
and is assumed to be heavy since it plays no role in the following.\footnote{
If we had defined the T action on the fermions 
$\Psi_1\stackrel{{\rm T}}{\longleftrightarrow}-\Sigma_0\Psi_2$, 
$\Psi_R\stackrel{{\rm T}}{\longrightarrow}-\Psi_R$
and the Yukawa Lagrangian with $\Omega$'s, 
$\lag_{Y_H} = -\kappa f 
\left(\overline\Psi_2\xi+ \overline\Psi_1\Sigma_0
\Omega\xi^\dagger\Omega\right)\Psi_R + {\rm h.c.}$, 
all new fermions would be T-odd and the new Lagrangian 
invariant under the new T-parity \cite{Low:2004xc}, 
but not under the full global symmetry because $\Omega$ does not 
commute with $SU(5)$ nor with $SO(5)$, though it does commute 
with the gauge group. 
Regardless, the explicit couplings entering in our 
calculation are the same in both cases.} 

The remaining leptons become massive through two other different 
Yukawa Lagrangians. 
The SM combination $l_L = \frac{l_{1L} - l_{2L}}{\sqrt 2}$ in Eqs. (\ref{Psimultiplets}--\ref{PsiT}) will obtain a mass through the following Yukawa couplings
\cite{Hubisz:2004ft,Chen:2006cs} 
\bea
{\cal L}_Y = \frac{i \lambda}{2 \sqrt 2} f \epsilon^{xyz}\epsilon^{rs} 
\left[ (\overline{\Psi_2^\chi})_x (\Sigma)_{ry} (\Sigma)_{sz} +  
(\overline{\Psi_1^{\tilde{\chi}}} \Sigma_0 \Omega)_x (\widetilde{\Sigma})_{ry} (\widetilde{\Sigma})_{sz} \right] 
\ell_R +{\rm h.c.}\ , 
\label{SMmasses}
\eea
where summation over $x, y, z = 3, 4, 5$ and $r, s = 1, 2$ is understood. 

Although all $SU(5)$ indices are contracted and the SM right-handed leptons are assumed to be 
$SU(5)$ singlets, this Lagrangian is not invariant under the $SU(5)$ global symmetry.
In addition to the breaking due to $\Sigma_0$ and the incomplete $\Psi_{1,2}$ multiplets, the 
$\Psi_{1,2}$ non-zero entries do not have the proper quantum numbers to construct the SM terms. 
Indeed, in order to enforce gauge invariance under the SM, extra $U(1)$ charges outside the 
$SU(5)$ are needed since $\ell_R$ is required to have hypercharge $-1$. 
Then, the term in brackets in Eq. (\ref{SMmasses}) must have hypercharge 1, but being an 
$SU(5)$ singlet this $U(1)$ charge must lie outside the $SU(5)$ as well.

Following Refs. \cite{Hubisz:2004ft,Chen:2006cs,Goto:2010sn} we compose two other 
incomplete $SU(5)$ multiplets in fundamental representations $\underline{5}$ and 
$\underline{5}^*$, respectively: 
\bea
\Psi^{\tilde{\chi}}_1=\left(\ba{c} \tilde{\chi} l_{1L} \\ 0 \\ 0 \ea\right),\quad
\Psi^\chi_2=\left(\ba{c} 0 \\ 0 \\ \chi l_{2L} \ea\right) , 
\label{Psichimultiplets}
\eea
where $\chi$ is a scalar with the proper charges to endow $\chi l_{2L}$ with the charges 
corresponding to the last two components of $\underline{5}^*$. 
More precisely, we will only require that it transforms properly under its gauged subgroup. 
The charge assignments fulfilling this requirement are gathered in Table \ref{chicharges} 
\cite{Goto:2010sn}. 
\begin{table}
\centering
\begin{tabular}{c||cccc}
 & \quad $Y'_1$ \quad & \quad $Y'_2$ \quad & \quad $Y''_1$ \quad & \quad $Y''_2$ \quad \T\B\\ 
\hline
$\chi$ & $\frac{2}{5}$ & $\frac{3}{5}$ & $-\frac{1}{2}$ &
$-\frac{1}{2}$ \T\B\\ 
$\tilde{\chi}$ & $\frac{3}{5}$ & $\frac{2}{5}$ & $-\frac{1}{2}$ & $-\frac{1}{2}$ \T\B\\ 
$l_{2L}$ & $-\frac{1}{5}$ & $-\frac{3}{10}$ & $0$ & $0$  \T\B\\
$l_{1L}$ & $-\frac{3}{10}$ & $-\frac{1}{5}$ & $0$ & $0$  \T\B\\
\hline
$\chi l_{2L}$ & $\frac{1}{5}$ & $\frac{3}{10}$ & $-\frac{1}{2}$ & $-\frac{1}{2}$  \T\B\\
$\tilde{\chi} l_{1L}$ & $\frac{3}{10}$ & $\frac{1}{5}$ & $-\frac{1}{2}$ & $-\frac{1}{2}$ \T\B \\
\end{tabular}
\caption{Charge assignment under $U(1)'_1 \times U(1)'_2 \times U(1)''_1 \times U(1)''_2$, 
where the first two single prime factors are the abelian subgroups inside $SU(5)$ while the 
double primed abelian groups live outside. 
The gauged $U(1)_1 \times U(1)_2$ correspond to the sums $Y_b = Y'_b + Y''_b,\ b = 1, 2$ 
where the total sum $Y = Y_1 + Y_2$ gives the hypercharge.}
\label{chicharges}
\end{table}
Thus, the introduction of $\chi$ allows us to change the sign of the gauged $U(1)$ charges in 
$SU(5)$ for $l_{2L}$ while also giving the correct hypercharge to $\chi l_{2L}$. 
The action under T-parity is then defined as
\bea
\Psi^{\tilde{\chi}}_1\stackrel{{\rm T}}{\longleftrightarrow}\Omega\Sigma_0\Psi^\chi_2\ .
\label{PsichiT}
\eea
An explicit realization of this extra scalar factor is obtained identifying $\chi$ 
with $(\Sigma^\dagger_{33})^{-\frac{1}{4}}$, and $\tilde{\chi}$ 
with 
$(\Sigma_{33})^{-\frac{1}{4}}$, which have the correct $Y_{1,2}$ charges, $(Y^\chi_1, Y^\chi_2) = 
(-\frac{1}{10}, \frac{1}{10})$ and 
$(Y^{\tilde\chi}_1, Y^{\tilde\chi}_2) = (\frac{1}{10}, -\frac{1}{10})$, and 
T-transformation properties \cite{Chen:2006cs}.\footnote{
This particular realization will not play an essential role in $h\to\tau\mu$ at the order to which we work.} 

Turning back to the partner lepton doublets $\tilde\psi_R$ in Eq. (\ref{complete}) they are T-odd, 
as desired, but must be heavy enough to agree with EWPD. 
Since they do not receive a mass from the Yukawa couplings in Eq. (\ref{mirror}) as the mirror 
leptons do, they remain massless as long as no other left-handed doublet is introduced to allow 
for the generation of a vector-like mass. 
The corresponding mechanism giving them a mass provides a new source of LFV if misaligned 
with the Yukawa couplings 
in Eqs. ({\ref{mirror}}) and (\ref{SMmasses}), 
making the discussion of their origin essential. 
Of course, as already emphasized and to be shown explicitly below, these partner leptons are 
also required to obtain a finite amplitude for $h\to\tau\mu$. 

A simple solution for giving $\tilde\psi_R$ a mass is to write an explicit mass term with an incomplete 
$SO(5)$ defining representation $\Psi_L = (\tilde\psi_L, 0, 0)^T$ 
which includes a left-handed counterpart $\tilde\psi_L = - i \sigma^2 (\tilde{l}_R)^c$ with
 which to form a vector like mass term 
\cite{Cheng:2004yc,Low:2004xc}.
There are other ways 
to generate a mass, but regardless of which mechanism is assumed it necessarily, but softly, breaks the 
$SO(5)$ to give a large mass to $\tilde\psi_R$ (and/or $\chi_R$) alone in the $\Psi_R$ multiplet.\footnote{
We could instead include the left-handed counterpart of  $\tilde\psi_R$ in an $SO(5)$ spinor (pseudo-real) representation $\underline{4}$, giving them a mass through a Yukawa coupling 
with new scalars also transforming as a $\underline{4}$, for 
$\underline{5} \subset \underline{4} \times \underline{4}$. 
In this way $\chi_R$ in $\Psi_R$ would also 
receive a mass when the SM neutral singlet within the scalar spinor representation gets a \emph{vev}. 
In contrast the SM neutral singlet lepton in the left-handed fermion multiplet $\underline{4}$ does not receive 
any mass 
because $\underline{5}$ is in the antisymmetric product of $\underline{4} \times \underline{4}$. 
(As already emphasized, the embedding of $SO(5) \subset SU(5)$ implied by the breaking along 
$\Sigma_0$ in Eq. (\ref{Sigma}) reduces the fundamental representation of $SU(5)$ to the defining (real)  representation of SO(5), $\underline{5} = \underline{5}$; and no pseudo-real representation is generated 
by tensor product of real ones. 
Hence, the \emph{vev} along the spinor representation of $SO(5)$ not only breaks this 
global symmetry group but its eventual embedding in an $SU(5)$ representation would be different to 
the assumed one in the non-linear realization of the LHT.)} 
This enters in corrections to the Higgs self-energy for which the $\tilde\psi_R$ mass serves as a finite 
cutoff at two loops~\cite{Cheng:2004yc,Low:2004xc}.
Thus these masses can not be taken arbitrarily large without reintroducing a fine tuning into the Higgs mass. 
As we show below, this soft breaking of $SO(5)$ also manifests itself at one loop and ${\cal O}(v^2/f^2)$ in 
$h\to\tau\mu$ as a logarithmic non-decoupling behavior when the partner lepton masses are taken large.
In our analysis we will simply parameterize the partner lepton masses as follows,
\bea
{\cal L}_M = - M^T \overline{\tilde\psi_L} \tilde\psi_R  +{\rm h.c.} = 
- M \overline{\tilde{l}_L} \tilde{l}_R +{\rm h.c.} \ , 
\label{psiRmasses}
\eea
where $\tilde\psi_R = - i \sigma^2 (\tilde{l}_L)^c$ and 
$\tilde{l}_L = (\tilde{\nu}_L \; \tilde{\ell}_L )^T$ (with $T$ meaning transpose).  

Although it does not have any consequence in our calculation, we comment that the Yukawa-type Lagrangian 
$\lag_{Y_H}$ fixes the transformation properties of the heavy fermions including their gauge couplings.
In particular the non-linear couplings of the right-handed heavy fermions \cite{Cheng:2004yc,Hubisz:2004ft} 
are fixed to be,
\bea
\lag_F&=&
i\overline\Psi_1\gamma^\mu D_\mu^*\Psi_1 \ + \ 
i\overline\Psi_2\gamma^\mu D_\mu\Psi_2  \nn
&+&i\overline{\Psi_R}\gamma^\mu\left(\partial_\mu+\frac{1}{2}\xi^\dagger 
(D_\mu\xi)+\frac{1}{2}\xi (\Sigma_0 D_\mu^* \Sigma_0 \xi^\dagger)\right)\Psi_R
\label{RHkineticterm} \\
&+& \Psi_R \rightarrow \Psi_L 
\nonumber
\eea
with the covariant derivative defined as
\bea
D_\mu&=&\partial_\mu-\sqrt{2}i g(W_{1\mu}^a Q_1^a+W_{2\mu}^a Q_2^a)
+\sqrt{2}i g'\left(Y_1 B_{1\mu}+Y_2 B_{2\mu}\right). 
\eea
This Lagrangian includes the proper ${\cal O}(v^2/f^2)$ 
couplings to Goldstone fields that render the one-loop lepton flavor changing amplitudes mediated by 
gauge bosons ultraviolet finite 
\cite{delAguila:2008zu,Goto:2008fj}. 
Finally, as discussed above, in order to assign the proper SM hypercharge $Y=-1$ to the charged 
right-handed leptons $\ell_R$, one can enlarge the global $SU(5)$ with two extra $U(1)$ groups 
for which we can write down the corresponding gauge and T-invariant Lagrangian 
\bea
\lag'_F=i\overline{\ell_R}\gamma^\mu (\partial_\mu+i g' Y B_\mu)\ell_R\ .
\eea

These are all the necessary Lagrangian terms 
for the lepton sector and the type of process we are interested in, 
up to family indices that we shall introduce in the following. 
In order to perform the calculation in the mass eigenstate basis we 
have to diagonalize the full Lagrangian 
\bea
\lag = 
\lag_{G} + \lag_{S} + \lag_{Y_H} + \lag_Y + \lag_M + \lag_{F}  + \lag_F' \ ,
\eea 
and re-express it in the mass eigenstate basis. 
The corresponding masses and eigenfields up to ${\cal O}(v^2/f^2)$ and the relevant Feynman rules are 
collected below and are obtained by
expanding $\lag$ to the required order.

\subsection{Mass eigenfields  and Feynman rules}
\label{Rules}

An important technical part of this paper is to prove unambiguously the finiteness of the process under 
study ($h \rightarrow \overline{\ell} \ell^\prime$
), and the need to include the full set of T-odd scalars 
and fermions. 
We will discuss the details of the calculation in next section, but first we collect here all the required 
Feynman rules and define the mass eigenstates.

\subsubsection{Mass eigenfields\label{apppf}}

{\it Gauge fields}. After EWSB the SM gauge boson 
mass eigenstates (see Eq. (\ref{GaugeTeven})), which are the T-even, are obtained by diagonalizing 
${\cal L}_S$ in Eq. (\ref{LS}): 
\bea
W^\pm=\frac{1}{\sqrt{2}}(W^1\mp i W^2)\ ,\quad
\left(\ba{c} Z \\ A \ea\right)=\left(\ba{cc} c_W & s_W \\ -s_W & c_W \ea\right)
\left(\ba{c} W^3 \\ B \ea\right) ,
\eea
with
\bea
W^a=\frac{W_1^a+W_2^a}{\sqrt{2}}\ ,\quad
B=\frac{B_1+B_2}{\sqrt{2}}\ ;
\eea
whereas the T-odd mass eigenstates, expanding the heavy sector up to order $v^2/f^2$, are
(see Eq. (\ref{GaugeTodd})): 
\bea
W_H^\pm=\frac{1}{\sqrt{2}}(W_H^1\mp i W_H^2)\ ,\quad
\left(\ba{c} Z_H \\ A_H \ea\right)=\left(\ba{cc} 1 & -x_H\dis\frac{v^2}{f^2} \\ x_H\dis\frac{v^2}{f^2} & 
1 \ea\right)\left(\ba{c} W_H^3 \\ B_H \ea\right),
\eea
with
\bea
W_H^a=\frac{W_1^a-W_2^a}{\sqrt{2}}\ ,\quad
B_H=\frac{B_1-B_2}{\sqrt{2}}\ ,\quad
x_H=\frac{5gg'}{4(5g^2-g'^2)}\ .
\label{XH}
\eea
Their masses to order $v^2/f^2$ are (with $e$ the electric charge, 
$s_W (c_W) = \sin \theta_W (\cos \theta_W)$, with $\theta_W$ the electroweak mixing angle, 
and $v\simeq246\mbox{ GeV}$)
\begin{align}
\label{MWZ}
&M_W=\frac{gv}{2}\left(1-\frac{v^2}{12f^2}\right)\ ,\quad
M_Z=M_W/c_W\ ,\quad
e=gs_W=g'c_W\ , &
\nonumber \\
&M_{W_H}=M_{Z_H}=gf\left(1-\frac{v^2}{8f^2}\right),\quad 
M_{A_H}=\frac{g'f}{\sqrt{5}}\left(1-\frac{5v^2}{8f^2}\right).&
\end{align}

\vskip 0.2cm
\noindent
{\it Scalar fields}. The scalar fields must be also rotated into the mass basis
\cite{Hubisz:2005tx}:
{\allowdisplaybreaks
\bea
\pi^0 &\to& \pi^0\left(1+\frac{v^2}{12f^2}\right)\ , \nonumber \\
\pi^\pm &\to& \pi^\pm\left(1+\frac{v^2}{12f^2}\right)\ , \nonumber \\
h &\to& h\ , \nonumber \\
\Phi^0 &\to& \Phi^0\left(1+\frac{v^2}{12f^2}\right)\ , \nonumber \\
\Phi^P &\to& \Phi^P + \left(\sqrt{10}\eta-\sqrt{2}\omega^0+
\Phi^P\right)\frac{v^2}{12f^2}\ , \nonumber \\
\Phi^\pm &\to& \Phi^\pm\left(1+\frac{v^2}{24f^2}\right)
\pm i\omega^\pm\frac{v^2}{12f^2}\ , \\
\Phi^{++} &\to& \Phi^{++}\ , \nonumber \\
\eta &\to& \eta + \frac{5g'\eta-4\sqrt{5}[g'(\omega^0+\sqrt{2}\Phi^P)
-6gx_H\omega^0]}{24g'}\frac{v^2}{f^2}\ , \nonumber \\
\omega^0 &\to& \omega^0 + \frac{5g(\omega^0+4\sqrt{2}\Phi^P)
-4\sqrt{5}\eta(5g+6g'x_H)}{120g}\frac{v^2}{f^2}\ , \nonumber \\
\omega^\pm &\to& \omega^\pm\left(1+\frac{v^2}{24f^2}\right)
\pm i\Phi^\pm\frac{v^2}{6f^2}\ . \nonumber
\eea}
The fields $\eta$, $\omega^0$ and $\omega^\pm$ are the Goldstone bosons of the gauge group 
$SU(2)_1 \times SU(2)_2 \times U(1)_1 \times U(1)_2$ breaking 
into the SM. 
They are eaten by the heavy gauge bosons 
$A_H$, $Z_H$ and $W_H^\pm$, respectively, 
while $\pi^0$ and $\pi^\pm$ are the Goldstone bosons of the SM 
gauge group. 
The physical pseudo-Goldstone bosons include the Higgs boson 
with mass $M_h \simeq 125$ GeV \cite{Olive:2016xmw}, 
and the scalar triplet of hypercharge 1, 
with a mass $M_{\Phi}$ which we assume to be a free parameter of order $f$ \cite{Han:2005nk}, 
though both masses are a priori calculable since they can be obtained in the LHT through a 
Coleman-Weinberg mechanism \cite{Coleman:1973jx}.

\vskip 0.2cm
\noindent
{\it Fermionic fields}. 
Fermion masses and mass eigenvectors are obtained from the diagonalization of 
the $3 \times 3$ matrices $\kappa$, $\lambda$ and $M$ in 
$\lag_{Y_H}$, $\lag_Y$ and $\lag_M$, respectively. 
For each of the three SM left-handed lepton doublets there is an extra
vector-like  
doublet, 
\bea
l_{rL}=\left(\ba{c} \nu_{rL} \\ \ell_{rL}\ea\right),\quad r=1,2\ , \quad
l_{HR}=\left(\ba{c} \nu_{HR} \\ \ell_{HR}\ea\right)\ ,
\eea
with (see Eqs. (\ref{Psimultiplets}-\ref{PsiRT}))
\bea
l_L=\frac{l_{1L}-l_{2L}}{\sqrt{2}}\ ,\quad
l_{HL}=\frac{l_{1L}+l_{2L}}{\sqrt{2}}\ ,\quad
l = \nu,\ell\ ,
\eea
where we have omitted flavor indices. 
The fields $\nu_L, \ell_L$ are the SM (T-even) left-handed leptons and $\nu_{H L},\ell_{H L}$ 
($\nu_{H R},\ell_{H R}$) are T-odd 
left (right) handed leptons with masses ${\cal O}(f)$.
In addition to three SM right-handed charged leptons $\ell_R$, 
which are assumed to be singlets under the non-abelian symmetries, there is a second heavy 
vector-like doublet per family, 
$\tilde{l}_{L} = ( \tilde{\nu}_{L} \;  \tilde{\ell}_{L} )^T$ and its Dirac partner 
$\tilde{l}_{R} = ( \tilde{\nu}_{R} \;  \tilde{\ell}_{R} )^T$.   
These come from the partner leptons $\tilde\psi_R$ needed to complete the $SO(5)$ multiplet 
$\Psi_R$ in Eqs. (\ref{complete}) and its partner $\tilde\psi_L$ in Eq. (\ref{psiRmasses}).

Let us now introduce flavor indices and masses and mass eigenleptons, which we shall denote 
as the current eigenfermions but with family indices.
Since T-parity is exact, the T-even SM charged leptons $\ell$ do not
mix with the heavy T-odd leptons.  
Thus the SM mass eigenstates result from the diagonalization of the $3
\times 3$ matrix $\lambda$  
in Eq. (\ref{SMmasses}) 
\bea
\lambda_{ij}\ v = (V^{\ell}_L)_{ik}\ m_{\ell_k}\ (V^{\ell\dagger}_R)_{kj}\ ,
\label{SMleptonmasses}
\eea
where $i, j, k = 1, 2, 3$, $V^{\ell}_{L, R}$ are $3 \times 3$ unitary
matrices and $m_{\ell_{1, 2, 3}} = m_{e, \mu, \tau}$ (up to
loop corrections $\mathcal{O}(v^2/f^2)$). 
The charged T-odd lepton masses result from diagonalizing 
a $6 \times 6$ matrix, which is block diagonal 
(see Eqs. (\ref{mirror}) and (\ref{psiRmasses})),
\bea
\begin{tabular}{lr}
$\left( \begin{array}{cc} \kappa_{ij} f \sqrt 2 & 0 \\ 0 & M_{i'j'} \end{array} \right) 
=$ & \\ 
\quad \quad $\left( \begin{array}{cc} (V^H_L)_{ik} & 0 \\ 0 & ({\tilde{V}}_L)_{i'k'} \end{array} \right) 
\left( \begin{array}{cc} m_{\ell_{H k}} & 0 \\ 0 & m_{\tilde{\ell}_{k'}} \end{array} \right)  
\left( \begin{array}{cc} (V^{H \dagger}_R)_{kj} & 0 \\ 0 & ({\tilde{V}}^{\dagger}_R)_{k'j'} \end{array} \right)$\ , 
& \\
\label{Toddleptonmasses}
\end{tabular}
\eea
where $i, j, k, i', j', k' = 1, 2, 3$ and $m_{\ell_{H 1, 2, 3},\ \tilde{\ell}_{1, 2, 3}}$ are the mass 
eigenvalues, and $V^{H}_{L, R}, \tilde{V}_{L, R}$ are $3 \times 3$ unitary matrices.\footnote{
The Yukawa Lagrangian in Eq. (\ref{mirror}) does not mix T-odd charged leptons without and 
with {\it tilde} because they are embedded with opposite (hyper)charges in $\Psi_R$. 
} 
The heavy (T-odd) charged and neutral leptons have the same mass and diagonalization matrices at leading order. 
At next order the $3 \times 3$ top-left block gets a correction 
$\sqrt 2 \kappa_{ij} f \rightarrow \sqrt 2 \kappa_{ij} f \left(1-\frac{v^2}{8f^2} \right)$ and hence, 
\bea
m_{\nu_{Hk}}=m_{\ell_{Hk}} \left(1-\frac{v^2}{8f^2}\right) \ .
\label{lHmasses}
\eea
There is also a correction of this order combining $\overline{\nu}_{H L}$ and $(\tilde{\nu}_{L})^c$ in which case a diagonalization of the corresponding $12 \times 12$ mass matrix for the heavy neutrinos is needed.
This gives corrections to heavy neutrino masses which are ${\cal O} (v^4/f^4)$ and mixings among heavy neutrinos 
${\cal O} (v^2/f^2)$.
This translates into higher order ${\cal O} (v^4/f^4)$ corrections to 
the one-loop Higgs decay under study and is therefore neglected. 

The misalignment between the mass matrices of the T-even (SM) and 
T-odd (heavy) leptons is a source of flavor mixing. 
We can work without loss of generality in the mass eigenstate basis 
diagonalizing the SM Yukawa couplings in ${\cal L}_Y$. 
We can then write the mixing matrices parameterizing the misalignment between the different leptonic sectors as 
\bea
V \equiv V_L^{H \dagger} V_L^\ell \ , \quad   W \equiv {\tilde{V}}_L^T V_R^H \ .
\label{mixingmatrices}
\eea
These will enter explicitly into the Feynman rules and amplitudes we study below.
As we will see, the new source of LFV we emphasize here enters through the mixing matrix $W$.

\subsubsection{Feynman rules\label{appfr}}

In order to calculate the Higgs decay into two different charged leptons at one loop, we need to complete the Feynman rules for the LHT worked out previously in the literature.\footnote{
In particular, in Ref. \cite{delAguila:2008zu} the Feynman rules in Appendix B.2 for one scalar and two right-handed leptons, $c_R$ in SFF, have a typo. 
The flavor subscript for $m_{\ell_i}$ should be $j$, and not $i$. 
The sign conventions are chosen to be compatible with those 
employed for the SM in Ref.~\cite{Denner:1991kt}, which coincide with those in 
Ref. \cite{Blanke:2006eb} 
up to a sign in the definition of the abelian gauge couplings in the covariant 
derivative in Eq. (\ref{derivative}).  
} 
In particular, we have to work out the terms of $\lag_S + \lag_{Y_H} + \lag_Y$ 
involving a Higgs up to order $v^2/f^2$. 
We just present the Feynman rules in the mass eigenstate basis 
which are necessary for the calculation of the LFV processes discussed in this work. 
They are given in Tables \ref{SFF}$-$\ref{SSV-SSS} in terms of generic couplings 
for the following general vertices involving 
scalars (S), fermions (F) and/or gauge bosons (V):
\bea
\label{couplingdefinition}
\mbox{[SFF]} &=& i (c_LP_L+c_RP_R)\ ,
\nonumber \\
\mbox{[SV$_\mu$V$_\nu$]} &=& i Kg^{\mu\nu}\ ,
\nonumber \\ 
\mbox{[V$_\mu$FF]} &=& i \gamma^\mu(g_LP_L+g_RP_R)\ ,
\nonumber \\
\mbox{[SSFF]} &=& i (f_LP_L+f_RP_R)\ ,
\\ 
\mbox{[S$(p_1)$S$(p_2)$V$_\mu$]} &=& i G (p_1-p_2)^\mu\ ,
\nonumber \\
\mbox{[SS$(p_1)$S$(p_2)$]} &=& i J \left( p_1^2 + p_2^2 + 4 p_1 \cdot p_2 \right)\ , \nonumber 
\eea
where all momenta are assumed incoming. 
The conjugate vertices are obtained replacing:
\bea
c_{L,R}\leftrightarrow c_{R,L}^*\ ,\; \; 
K\leftrightarrow K^*\ ,\; \; 
g_{L,R}\leftrightarrow g_{L,R}^*\ ,\; \; 
f_{L,R}\leftrightarrow f_{R,L}^*\ ,\; \; 
G\leftrightarrow G^*\ ,\; \; 
J\leftrightarrow J^*\ .
\eea
\begin{table}
	\begin{center}
		\begin{tabular}{c||c|c}
			[SFF] & $c_L$ & $c_R$ \\
			\hline
			$h\ \overline{\ell_i} \ell_j $ & $- \delta_{ij} \frac{m_{\ell_j}}{v} \left( 1 - \frac{v^2}{6f^2} \right)$ & 
			$- \delta_{ij} \frac{m_{\ell_j}}{v} \left( 1 - \frac{v^2}{6f^2} \right)$ \\
			$h\ \overline{\nu_{H i}}\ \nu_{H j} $ & $\frac{m_{\ell_{H i}}}{v} \delta_{ij} \frac{v^2}{4f^2}$ & 
			$\frac{m_{\ell_{H i}}}{v} \delta_{ij} \frac{v^2}{4f^2}$ \\
			$h\ \overline{\ell_{H i}}\ \ell_{H j} $ & $0$ & $0$ \\
			$h\ \overline{\nu_{H i}}\ \tilde{\nu}_{j}^c $ & $0$ & 
			$\frac{m_{\ell_{H i}}}{v} W^\dagger_{ij} \frac{v^2}{4f^2}$ \\
			$\Phi^0 \ \overline{\ell_{H i}}\ \ell_{j} $ & $0$ & 
			$V_{ij} \frac{m_{\ell_j}}{\sqrt 2 f} \left( 1 + \frac{v^2}{4f^2} \right)$ \\
			$\Phi^P \ \overline{\ell_{H i}}\ \ell_{j} $ & $0$ & 
			$i V_{ij} \frac{m_{\ell_j}}{\sqrt 2 f} \left( 1 + \frac{v^2}{4f^2} \right)$ \\
			$\Phi^+\ \overline{\nu_{H i}}\ \ell_{j} $ & $\frac{m_{\ell_{H i}}}{\sqrt 2 f} V_{ij} \frac{v^2}{8 f^2}$ & 
			$V_{ij} \frac{m_{\ell_j}}{\sqrt 2 f} \left( 1 - \frac{v^2}{8f^2} \right)$ \\
			$\Phi^+\ \overline{\tilde{\nu}^c_{i}}\ \ell_{j} $ & 
			$W_{ik} \frac{m_{\ell_{H k}}}{\sqrt 2 f} V_{kj}$ 
			& $0$ \\
			$\Phi^{++}\ \overline{\tilde{\ell}_i^c}\ \ell_j$ & 
			$- W_{ik} \frac{m_{\ell_{H k}}}{f} V_{kj}$ 
			& $0$ \\
			$\eta \ \overline{\ell_{H i}}\ \ell_{j} $ & 
			$i \frac{m_{\ell_{H i}}}{2 \sqrt 5 f} V_{ij} \left[ 1 - (\frac{5}{8} + x_H t_W) \frac{v^2}{f^2} \right]$ & 
			$-i V_{ij} \frac{m_{\ell_j}}{2 \sqrt 5 f} \left[ 1 - (\frac{5}{8} + x_H t_W) \frac{v^2}{f^2} \right]$ \\
			$\omega^0\ \overline{\ell_{H i}}\ \ell_{j} $ & 
			$i \frac{m_{\ell_{H i}}}{2 f} V_{ij} \left[ 1 - (\frac{1}{8} - \frac{x_H}{t_W}) \frac{v^2}{f^2} \right]$ & 
			$-i V_{ij} \frac{m_{\ell_j}}{2 f} \left[ 1 - (\frac{1}{8} - \frac{x_H}{t_W}) \frac{v^2}{f^2} \right]$ \\
			$\omega^+ \ \overline{\nu_{H i}}\ \ell_{j} $ & 
			$-i \frac{m_{\ell_{H i}}}{\sqrt 2 f} V_{ij}$ 
			& $i V_{ij} \frac{m_{\ell_j}}{\sqrt 2 f} \left( 1 + \frac{v^2}{8 f^2} \right)$ \\
			$\omega^+ \ \overline{\tilde{\nu}^c_{i}}\ \ell_{j} $ & 
			$i W_{ik} \frac{m_{\ell_{H k}}}{\sqrt 2 f} V_{kj} \frac{v^2}{8 f^2}$ & $0$ \\
		\end{tabular}
	\end{center}
	\caption{Scalar-Fermion-Fermion couplings at ${\cal O} ( v^2 / f^2 )$. 
		We use $t_W = \tan \theta_W = \frac{s_W}{c_W}$. 
		(The coupling $\Phi^{++} \overline{\tilde{\ell}_i^c} \ell_j$ does not enter in the calculation 
		of $h \rightarrow \overline{\ell} \ell'$, but it does in 
		$Z, \gamma \rightarrow \overline{\ell} \ell'$ in order to cancel other contributions.)}
	\label{SFF}
\end{table}
\begin{table}
\centering
\begin{tabular}{c||ccc||c}
[SV$_\mu$V$_\nu$] & $K$ & $\quad$ & [V$_\mu$FF] & $g_L$ \\
\cline{1-2} \cline{4-5}
$h\ W^+_H\ W^-_H$ & $- g^2 \frac{v}{2}$ && 
$W_H^+\ \overline{\nu_{H i}}\ \ell_j $ & $\frac{g}{\sqrt 2} V_{ij}$ \\
$h\ Z_H\ Z_H $ & $- g^2 \frac{v}{4}$ && 
$Z_H\ \overline{\ell_{H i}} \ell_j $ & $- \left( \frac{g}{2} + \frac{g^\prime}{10} x_H \frac{v^2}{f^2} \right) V_{ij}$ \\
$h\ A_H\ A_H $ & $- g^{\prime2} \frac{v}{4}$ && 
$A_H\ \overline{\ell_{H i}} \ell_j $ & $ \left( \frac{g^\prime}{10} - \frac{g}{2} x_H \frac{v^2}{f^2} \right) V_{ij}$ \\
$h\ Z_H\ A_H $ & $- g g^{\prime} \frac{v}{4}$ &&
\multicolumn{2}{c}{} \\
\end{tabular}
\caption{Scalar-Vector-Vector and Vector-Fermion-Fermion couplings at ${\cal O} (v^2/f^2)$. 
The right-handed Vector-Fermion-Fermion couplings $g_R$ vanish.}
\label{SVV-VFF}
\end{table}
\begin{table}
	\begin{center}
		\begin{tabular}{c||c|c}
			[SSFF] & $f_L$ & $f_R$ \\
			\hline
			$h\ \Phi^0\ \overline{\ell_{H i}}\ \ell_j $ & $0$ & 
			$V_{ij} \frac{m_{\ell_j}}{\sqrt 2 v f} \left( 1 + \frac{5v^2}{12f^2} \right)$ \\
			$h\ \Phi^P\ \overline{\ell_{H i}}\ \ell_j $ & $0$ & 
			$i V_{ij} \frac{m_{\ell_j}}{\sqrt 2 v f} \left( 1 + \frac{5v^2}{12f^2} \right)$ \\
			$h\ \Phi^+\ \overline{\nu_{H i}}\ \ell_j $ & $- \frac{m_{\ell_{H i}}}{\sqrt 2 v f} V_{ij} \frac{v^2}{12f^2} $ & 
			$V_{ij} \frac{m_{\ell_j}}{\sqrt 2 v f} \left( 1 - \frac{7v^2}{24f^2} \right)$ \\
			$h\ \Phi^+\ \overline{\tilde{\nu}^c_{i}}\ \ell_j $ & 
			$- W_{ik} \frac{m_{\ell_{H k}}}{\sqrt 2 v f} V_{kj} \frac{v^2}{12f^2} $  & 
			$0$ \\
			$h\ \eta\ \overline{\ell_{H i}}\ \ell_j $ & $0$ & 
			$-i V_{ij} \frac{m_{\ell_j}}{2 \sqrt 5 v f} \left[ 1 + (\frac{7}{8} - x_H \frac{s_W}{c_W}) \frac{v^2}{f^2} \right]$ \\
			$h\ \omega^0\ \overline{\ell_{H i}}\ \ell_j $ & $0$ & 
			$-i V_{ij} \frac{m_{\ell_j}}{2 v f} \left[ 1 - (\frac{5}{8} - x_H \frac{c_W}{s_W}) \frac{v^2}{f^2} \right]$ \\
			$h\ \omega^+\ \overline{\nu_{H i}}\ \ell_j $ & $i \frac{m_{\ell_{H i}}}{\sqrt 2 v f} V_{ij} \frac{v^2}{12f^2} $ & 
			$i V_{ij} \frac{m_{\ell_j}}{\sqrt 2 v f} \left( 1 - \frac{v^2}{24f^2} \right)$ \\
			$h\ \omega^+\ \overline{\tilde{\nu}^c_{i}}\ \ell_j $ & 
			$i W_{ik} \frac{m_{\ell_{H k}}}{\sqrt 2 v f} V_{kj} \frac{v^2}{12f^2} $ & 
			$0$ \\
		\end{tabular}
	\end{center}
	\caption{Scalar-Scalar-Fermion-Fermion couplings at ${\cal O} ( v^2 / f^2 )$. 
		The coupling $h \phi^{++} \overline{\tilde{\ell}_i^c} \ell_j$ does vanish.}
	\label{SSFF}
\end{table}
\begin{table}
	\begin{center}
		\begin{tabular}{c||ccc||c}
			[S$(p_1)$S$(p_2)$V$_\mu$] & $G$ && [SS$(p_1)$S$(p_2)$] & $J$ \\
			\cline{1-2} \cline{4-5}
			$h\ \Phi^+\ W_H^-$ & $g \frac{v}{4 f}$ && 
			$h\ \Phi^0\ \Phi^0$ & $\frac{v}{6 f^2}$ \\
			$h\ \eta\ A_H^-$ & $i {\sqrt 5} g' \frac{v}{4 f}$ &&
			$h\ \Phi^P\ \Phi^P$ & $\frac{v}{6 f^2}$ \\
			$h\ \omega^0\ Z_H^-$ & $- i g \frac{v}{4 f}$ &&
			$h\ \Phi^P\ \eta$ & $-{\sqrt{\frac{5}{2}}} \frac{v}{6 f^2}$ \\
			$h\ \omega^+\ W_H^-$ & $- i g \frac{v}{4 f}$ &&
			$h\ \Phi^P\ \omega^0$ & $\frac{1}{\sqrt 2} \frac{v}{6 f^2}$ \\
			\multicolumn{2}{c}{} && $h\ \Phi^+\ \Phi^-$ & $\frac{v}{12 f^2}$ \\
			\multicolumn{2}{c}{} && $h\ \Phi^+\ \omega^-$ & $i \frac{v}{12 f^2}$ \\
			\multicolumn{2}{c}{} && $h\ \eta\ \eta$ & $\frac{5 v}{12 f^2}$ \\
			\multicolumn{2}{c}{} && $h\ \eta\ \omega^0$ & $- \frac{\sqrt 5 v}{12 f^2}$ \\
			\multicolumn{2}{c}{} && $h\ \omega^0\ \omega^0$ & $\frac{v}{12 f^2}$ \\
			\multicolumn{2}{c}{} && $h\ \omega^+\ \omega^-$ & $\frac{v}{12 f^2}$ \\
		\end{tabular}
	\end{center}
	\caption{Scalar-Scalar-Vector and Scalar-Scalar-Scalar couplings at ${\cal O} ( v^2 / f^2 )$. 
		Other combinations also involving $h$ vanish.}
	\label{SSV-SSS}
\end{table}
%

\section{Higgs coupling to a pair of different charged leptons at one loop in the LHT}
\label{Calculation}

The global symmetries of the LHT prevent tree level LFV
Higgs decays, but they are generated at one loop via the T-odd particles. 
Since they are forbidden at tree level, the one loop decays are finite and  a prediction of the LHT as we explicitly show below. 
To gain intuition for the parametric dependence, first there is the universal one loop-factor $(16 \pi ^2)^{-1}$. 
In addition, since we can always assume without loss of generality that the SM charged lepton Yukawa couplings are diagonal, any LFV Higgs decays must proceed via higher dimensional operators which are suppressed by the scale of new physics $f$ and scale like $v^2/f^2$. 
Moreover, the LFV Higgs decays must be proportional to the SM Yukawa couplings $\lambda$ and to the heavy lepton source of flavor violation $\Delta \kappa^2 \sin 2\theta$ 
because it is the misalignment of both sets of couplings which leads to LFV in the Higgs 
decay.\footnote{The Yukawa couplings $\kappa$ giving large masses to T-odd particles must enter 
squared because only T-odd particles run in the loop. In fact, it is their difference 
$\Delta\kappa^2$ that enters, 
whereas $\sin 2\theta$ parameterizes the misalignment.}
Taking this into account and neglecting integral functions of mass ratios of heavy particles which are ${\cal O} (1)$ for masses not much larger than $f$,\footnote{As we discuss below, for very large masses of the partner lepton doublets these finite integrals can manifest a logarithmic behavior.}
the amplitude for the Higgs decay $h \rightarrow \overline{\ell} \ell^\prime$ scales as 
\bea
{\mathcal M} \propto \frac{1}{16 \pi ^2} \frac{v^2}{f^2}\ \lambda\ \Delta \kappa^2 \sin 2\theta \ .  
\eea 

The two main results of this paper are then first to prove the finiteness of one-loop $h \rightarrow \overline{\ell} \ell^\prime$ decays in the LHT. 
As we show below, this relies on non-trivial cancellations among contributions from the heavy mirror and partner T-odd leptons. Second, the identification of new sources of LFV present in this class of models which contribute to LFV Higgs decays, as well as in general to all LFV amplitudes in the LHT.

\subsection{One-loop contribution of T-odd particles to $h \rightarrow \overline{\ell} \ell^\prime$ in the LHT}
\label{Finiteness}

\begin{figure}
  \centering
  \begin{tabular}{cccc}
\includegraphics[scale=0.65]{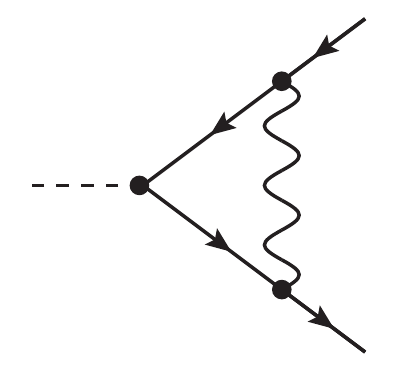} &
\includegraphics[scale=0.65]{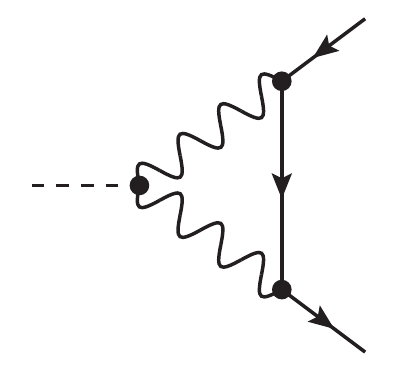} &
\includegraphics[scale=0.65]{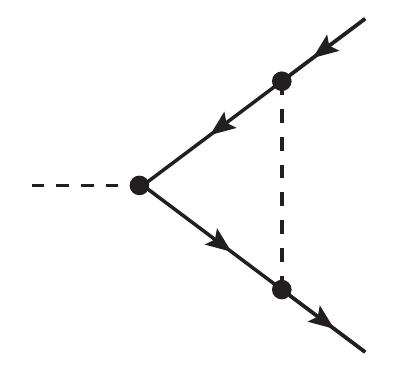} &
\includegraphics[scale=0.65]{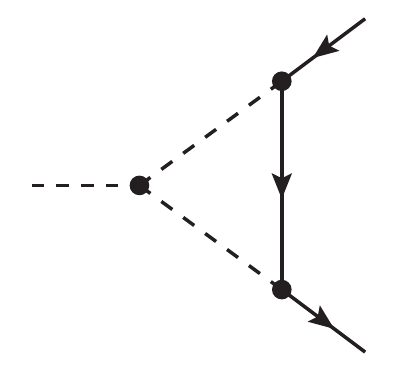} \\
I & II & III & IV \\
\includegraphics[scale=0.65]{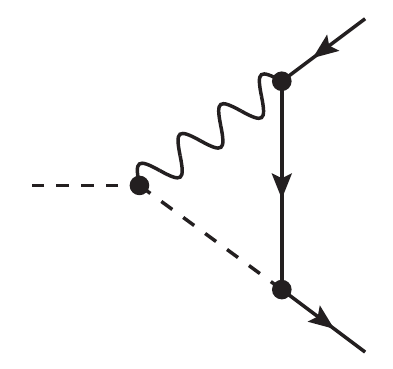} &
\includegraphics[scale=0.65]{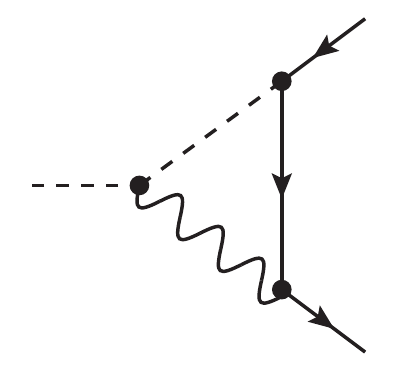} &
\includegraphics[scale=0.65]{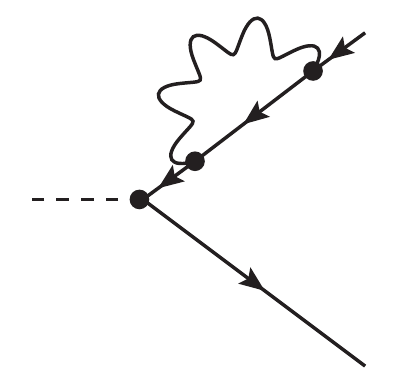} &
\includegraphics[scale=0.65]{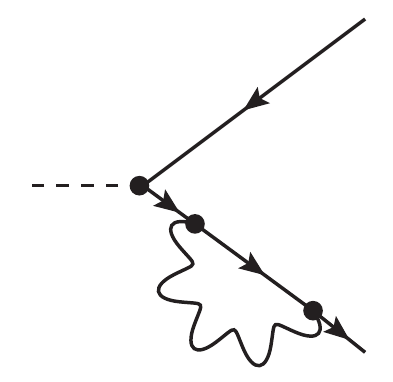} \\
V & VI & VII & VIII \\
\includegraphics[scale=0.65]{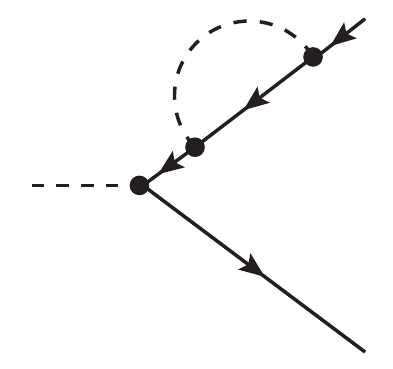} &
\includegraphics[scale=0.65]{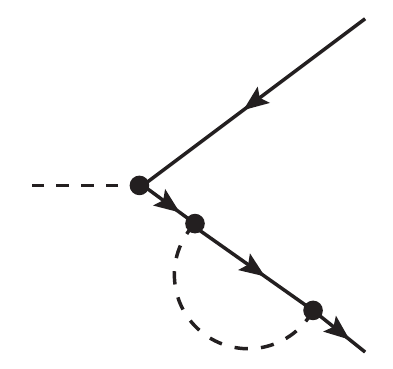} &
\includegraphics[scale=0.65]{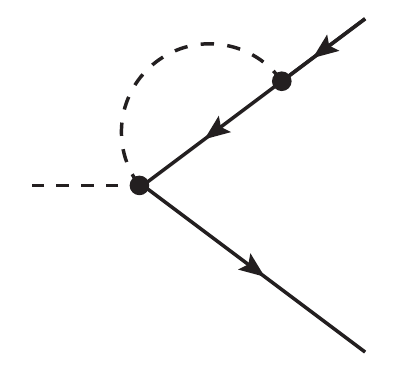} &
\includegraphics[scale=0.65]{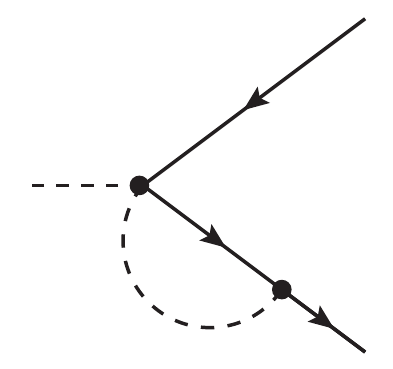} \\
IX & X & XI & XII
\end{tabular}
\caption{Topologies contributing to $h \rightarrow \overline{\ell} \ell'$.}
\label{diagrams}
\end{figure}

We work in the renormalizable 't Hooft-Feynman gauge. 
This process shows significant differences with the corresponding gauge boson mediated processes 
$Z, \gamma  \rightarrow \overline{\ell} \ell^\prime$. 
First, its finiteness requires the exchange of the full set of T-odd particles 
in the scalar and lepton sectors of the model introduced in the former section. 
This is apparent from inspection of the divergent contributions to the $h \rightarrow \overline{\ell} \ell^\prime$ amplitude. 
The different topologies are depicted in Figure \ref{diagrams}.\footnote{
There are new topologies with non-renormalizable 
couplings in this case, compared with the corresponding 
gauge transitions \cite{delAguila:2008zu}.
}
Individually these amplitudes produce infinite and finite pieces at ${\cal O} (1)$. 
This means that they are suppressed only by the one-loop factor as well as Yukawa couplings and mixing angles, but not $v^2/f^2$. 
The total sum however cancels as it must since there is no available counterterm. 
Using the Feynman rules in the previous section one 
finds the divergent contributions listed in Table \ref{1infinities}.\footnote{
As shown in next section, the divergent part of the amplitude can be written  ${\cal M}_{\rm div} (h \rightarrow \overline{\ell}\ell') = {1\over 16\pi^2} (C^{(1)}_{\rm UV} + {v^2 \over f^2} C_{\rm UV}^{(\frac{v^2}{f^2})}) \frac{1}{\epsilon} 
\sum_{i=1}^3 V^\dagger_{\ell' i} V_{i \ell} \frac{m_{\ell_{H i}}^2}{f^2}
\bar{u}(p^\prime,m_{\ell'}) \left( \frac{m_{\ell'}}{v} P_L + \frac{m_\ell}{v} P_R \right)  v(p,m_\ell)$.  
}
\begin{table}
\centering
\begin{tabular}{r||cccccccc|c}
$C_{\rm UV}^{(1)}$ & I & II & III& 
 IV& V+VI& VII+VIII & IX+X & 
 XI+XII &\ Sum \B\\
\hline
$\omega,\nu_H$ & -- & -- & $\bullet$ & $\bullet$ & -- & -- & $1$ &
$-1$ & $\bullet$ \TT\B
\\
$\omega^0,\ell_H$ & -- & -- & $\bullet$ & $\bullet$ & -- & -- & $\frac{1}{2}$ & $-\frac{1}{2}$ & $\bullet$ \TT\B\\
$\eta,\ell_H$ & -- & -- & $\bullet$ & $\bullet$ & -- & -- & $\frac{1}{10}$ & $-\frac{1}{10}$ & $\bullet$ \TT\B\\
\hline
Total & -- & -- & $\bullet$ & $\bullet$ & -- & -- & $\frac{8}{5}$ & $-\frac{8}{5}$ & $\bullet$ \TT\\
\end{tabular}
\caption{Divergent contributions proportional to $\frac{1}{\epsilon}$, with $\epsilon = 4 - d$ the extra 
dimensions in dimensional regularization, of each particle set running in the loop and topology in 
Figure \ref{diagrams} contributing at ${\cal O} (1)$. 
A dash means that the field set does not run in the diagram, whereas a dot indicates that the infinite and finite parts vanish.}
\label{1infinities}
\end{table}
The numbers $C^{(1)}_{\rm UV}$ are the coefficients, up to a global factor, of $\frac{1}{\epsilon}$, 
with $\epsilon = 4 - d$ the extra dimensions in dimensional regularization. 
The dashes mean 
that the fields in the row do not close the loop of the topology in the column. 
The dots stand for the vanishing of the infinite and finite pieces 
of the corresponding diagrams. 
As indicated by the bullets, the sums of the different topologies in the last column give not only finite but vanishing contributions (the sum of the contributions with the same topology gathered in the last row are non-zero and infinite in general, but their total sum does cancel). 

The ${\cal O} (v^2/f^2)$ contributions are more interesting. 
Again there is no counterterm for the corresponding operator of dimension 6 indicating that amplitude must be finite which we check explicitly. 
In Table \ref{vfinfinities} we gather the coefficients $C_{\rm UV}^{(\frac{v^2}{f^2})}$ of the divergent pieces for the different field contributions (rows) to a corresponding topology (columns). 
\begin{table}
	\begin{center}
		\hspace*{-0.65cm}
		\begin{tabular}{r||cccccccc|c}
			$C_{\rm UV}^{(\frac{v^2}{f^2})}$ & I & II & III & 
			 IV & V+VI & VII+VIII & IX+X & 
			 XI+XII &\ Sum \B\\
			\hline
			$W_H,\nu_H$ & 0 & 0 & -- & -- & -- & $\bullet$ & -- & -- & 0 \TT\B\\ 
			$W_H,\omega,\nu_H$ & -- & -- & -- & -- & 0 & -- & -- & -- & 0 \TT\B\\ 
			$\omega,\nu_H$ & -- & -- & $\frac{1}{4}$ & $-\frac{1}{8}$ & -- & -- & $-\frac{1}{6}$ & $\frac{5}{24}$ & $\frac{1}{6}$ \\ 
			\hline
			$Z_H,\ell_H$ & $\bullet$ & 0 & -- & -- & -- & $\bullet$ & -- & -- & 0 \TT\B\\ 
			$Z_H,\omega^0,\ell_H$ & -- & -- & -- & -- & 0 & -- & -- & -- & 0 \TT\B\\
			$\omega^0,\ell_H$ & -- & -- & $\bullet$ & $-\frac{1}{16}$ & -- & -- & 
			$-\frac{5}{24}+x_H\frac{c_W}{s_W}$ & $\frac{3}{8}-x_H\frac{c_W}{s_W}$ & $\frac{5}{48}$ \TT\B\\
			\hline
			$A_H\ell_H$ & $\bullet$ & 0 & -- & -- & -- & $\bullet$ & -- & -- & 0 \TT\B\\
			$A_H,\eta,\ell_H$ & -- & -- & -- & -- & 0 & -- & -- & -- & 0 \TT\B\\
			$\eta,\ell_H$ & -- & -- & $\bullet$ & $-\frac{1}{16}$ & -- & -- & 
			$-\frac{17}{120}-x_H\frac{s_W}{5c_W}$ & $-\frac{1}{40}+x_H\frac{s_W}{5c_W}$ & $-\frac{11}{48}$ \TT\B\\
			\hline
			$Z_H,A_H,\ell_H$ & -- & 0 & -- & -- & -- & -- & -- & -- & 0 \TT\B\\
			$\omega^0,\eta,\ell_H$ & -- & -- & -- & $\frac{1}{8}$ & -- & -- & -- & -- & $\frac{1}{8}$ \TT\B\\
			\hline
			$W_H,\Phi,\nu_H$ & -- & -- & -- & -- & 0 & -- & -- & -- & 0 \TT\B\\ 
			$\Phi,\nu_H$ & -- & -- & $\bullet$ & $\bullet$ & -- & -- & $-\frac{1}{8}$ & $\frac{1}{24}$ & $-\frac{1}{12}$ \TT\B\\ 
			\hline
			$\omega,\Phi,\nu_H$ & -- & -- & -- & $\frac{1}{6}$ & -- & -- & -- & -- & $\frac{1}{6}$ \TT\B\\ 
			\hline
			$\omega^0,\Phi^P,\ell_H$ & -- & -- & -- & $\frac{1}{24}$ & -- & -- & -- & -- & $\frac{1}{24}$ \TT\B\\
			\hline
			$\eta,\Phi^P,\ell_H$ & -- & -- & -- & $-\frac{1}{24}$ & -- & -- & -- & -- & $-\frac{1}{24}$ \TT\B\\
			\hline
			$\Phi,\tilde\nu^c$ & -- & -- & $-\frac{1}{4}$ & $\frac{1}{24}$ & -- & -- & $\bullet$ & $-\frac{1}{24}$ & $-\frac{1}{4}$ \TT\B\\
			%
			\hline
			Total & 0 & 0 & 0 & $\frac{1}{12}$ & 0  & $\bullet$ & $-\frac{47}{120}$ & $\frac{37}{120}$& 0 \TT\\
		\end{tabular}
		\caption{As in Table 6 but to ${\cal O} ( v^2 / f^2 )$. 
			$x_H = \frac{5t_W}{4(5-t_W^2)}$ is defined in Eq. (2.32) with $t_W = \frac{s_W}{c_W}$.}
		\label{vfinfinities}
	\end{center}
\end{table}
The notation is as in Table \ref{1infinities}, but now a 0 means that only the infinite piece cancels. 
As can be observed by summing the entries of the last column, which adds to zero, the contribution of the charged diagrams (those exchanging heavy neutrinos) is finite. 
It is clear however from examining the table, and looking for example at the $\omega, \nu_H$ and 
$\Phi, \tilde\nu^c, \nu_H$ contributions to the III topology, that the scalar triplet of hypercharge 1 as well 
as the additional vector-like partner lepton doublets have to be taken into account to obtain a finite result. 
This was perhaps expected since the electroweak triplets and partner leptons are needed in order to guarantee that the $SO(5)$ global symmetry is preserved.
Quite often however it has been assumed that they are heavy enough to be ignored, that they can be decoupled. 
This is true for gauge boson mediated LFV processes where chiral symmetry allows for the $\Phi$ and $\tilde{\nu}^c$ contributions to be decoupled without introducing a divergence, but clearly does not hold for $h\to\tau\mu$. 
We will further discuss the $Z, \gamma  \rightarrow \overline{\ell} \ell^\prime$ amplitudes elsewhere. 

\subsection{New sources of flavor violation contributing to $h \rightarrow \overline{\ell} \ell^\prime$ in the LHT}
\label{Sources}

We work out the finite part in this subsection to discuss the sources 
of LFV in the LHT, and their behavior in the non-decoupling limit.\footnote{
We have validated our calculation using the Mathematica 
package FeynCalc \cite{Mertig:1990an,Shtabovenko:2016sxi}.
}
These sources also contribute to other processes like $Z, \gamma \rightarrow \overline{\ell} \ell'$, already discussed in the literature  
\cite{Blanke:2006eb,delAguila:2008zu}, though only the approximation of decoupled partner leptons and small momenta. 
These assumptions can not be made in general and in particular for Higgs decays. 
First, as shown in the previous subsection, the partner leptons are needed to make $h \rightarrow \overline{\ell} \ell^\prime$ finite. 
The fact that the partner lepton mass, which introduces a soft breaking of the global $SO(5)$, can not be taken to infinity and decoupled manifests as a logarithmic dependence of the $h \rightarrow \overline{\ell} \ell^\prime$ amplitude on its mass, as shown below. 
Second, for on and off shell Higgs and $Z$ decays at the LHC there will be significant transfer of momenta. 
Thus a global fit using the complete expressions for the LFV gauge boson mediated processes is also necessary \cite{inpreparation}, but we focus here first on the $h\to\tau\mu$ decay. 

Using the Feynman rules in the previous section, we calculate the one-loop contributions of the T-odd gauge bosons, scalars and fermions to the $h \rightarrow \overline{\ell} \ell^\prime$ amplitude 
in the LHT.
This can be written as
\begin{eqnarray}
\label{amplitude}
{{\cal M}}(h \rightarrow \overline{\ell}\ell') = 
\bar{u}(p^\prime,m_{\ell'})  \left( \frac{m_{\ell'}}{v} c^{\ell \ell'}_L P_L + 
\frac{m_\ell}{v} c^{\ell \ell'}_R P_R \right)  v(p,m_\ell)\ , 
\label{finiteamplitude}
\end{eqnarray}
with 
\begin{eqnarray}
\begin{aligned}
c^{\ell \ell'}_{L(R)} = 
\frac{\alpha_W}{4 \pi} \frac{v^2}{f^2} \left[ \sum_{i=1}^3 \right. 
& V^\dagger_{\ell' i} V_{i \ell}\ F (m_{\ell_{H i}}, M_{W_H}, M_{A_H}, M_\Phi) \\ 
\sum_{i,j,k=1}^{3} &\left. V^\dagger_{\ell' i} \frac{m_{\ell_{H i}}}{M_{W_H}} W^\dagger_{i j} W_{j k} 
\frac{m_{\ell_{H k}}}{M_{W_H}} V_{k \ell}\ G (m_{\tilde{\nu}^c_j}, m_{\ell_{H k(i)}}, M_\Phi) \right]\ . 
\label{AL-AR}
\end{aligned}
\end{eqnarray}
Note the different index contraction for $c^{\ell \ell'}_L$ and $c^{\ell \ell'}_R$, and the mass relations in Eqs. (\ref{MWZ}) and (\ref{lHmasses}) between $M_{W_H}$ and $M_{Z_H}$ and between $m_{\ell_{H i}}$ and $m_{\nu_{H i}}$, respectively.
The function $F$ is more involved and worked out in the Appendix. Making use of the scalar integrals of three-point functions in Ref. \cite{Passarino:1978jh} (we omit the first three arguments of the three-point functions, which we take equal to $0$ 
for the external fermion momenta and to $Q^2 = M_h^2$ for the Higgs momentum), 
the function $G$ in the second term reduces to   
\begin{eqnarray}
\begin{aligned}
G (m_{\tilde{\nu}^c_j}, m_{\ell_{H k}}, M_\Phi) & 
= \frac{1}{16} - \frac{1}{2} C_{00} (M_\Phi^2, m_{\tilde{\nu}^c_j}^2, m_{\ell_{H k}}^2) \\ 
& - \frac{1}{8} m_{\tilde{\nu}^c_j} m_{\ell_{H k}} C_{0} (M_\Phi^2, m_{\tilde{\nu}^c_j}^2, m_{\ell_{H k}}^2) 
 - \frac{1}{12} M_\Phi^2 C_{1} (m_{\tilde{\nu}^c_j}^2, M_\Phi^2, M_\Phi^2) \ . 
\label{G}
\end{aligned}
\end{eqnarray}
As is apparent from Eq. (\ref{AL-AR}), there are two sources 
of LFV present in the LHT model. 
They are proportional to the mixing matrices $V$ and $W$ in Eq. (\ref{mixingmatrices}) parameterizing the misalignment of the light lepton doublets with the heavy (vector-like) ones. 
The unitary matrix $V$ describes the flavor rotation between the former and their T-odd (left-handed) doublet (mirror) partners.
The unitary matrix $W$ parameterizes flavor rotations between the right-handed doublet counterparts of the latter and the extra doublets required to give the partner leptons a mass. 

The two sources of LFV in Eq. (\ref{AL-AR}) have a different 
dependence on the masses of the heavy vector-like leptons. 
The first source coming from the misalignment of the mirror leptons and involving only the rotation matrix $V$ depend on the mirror lepton masses, but not on the partner lepton masses, and only decouples when the scale $f$ is taken large.
In contrast, the second source of LFV resulting from the misalignment 
of the additional lepton doublets and involving both rotation matrices $V$ and $W$ is also function of the extra partner lepton masses $m_{\tilde{l}_ i}$, which have a different origin and may be a priori much larger than the scale $f$. In this limit this contribution grows logarithmically, reflecting the need for further NP to provide these leptons a mass. 
The behavior in this limit is dictated by the second term in Eq. (\ref{G}) 
proportional to $C_{00}$ and coming from topology III with particle content $\Phi, \tilde{\nu}^c, \nu_H$. 
This term grows as $\ln m_{\tilde{\nu}^c_j}$ and hence, there is no decoupling. 

\section{Model dependent limits on the Higgs coupling to a pair of different leptons in the LHT}
\label{Phenomenology}

We now estimate the corresponding branching ratio for the most
interesting experimental channel $h\to \tau^+ \mu^- + \tau^- \mu^+$.
To this end we still have to correct for the
final mass eigenstates. Indeed, the contributions
to $h \rightarrow {\overline{\ell}} \ell'$ in Eq.
(\ref{finiteamplitude}) imply that the corresponding off-diagonal
entries $\ell \ell'$ of the charged
lepton mass matrix also receive one-loop
corrections with the Higgs field insertion replaced by
the {\it vev}. Hence, a further diagonalization of this mass matrix is
required to obtain the lepton mass eigenstates
at the order in which we work. This diagonalization and
correction has been discussed in Ref. \cite{delAguila:2000aa}
for the quark sector (see also Ref. \cite{Agashe:2009di}),
and amounts to an extra multiplicative factor 2/3
for the amplitude in Eq. (\ref{finiteamplitude}) for the
actual final lepton mass eigenstates.\footnote{In contrast, for the
  gauge couplings the effect of this final rotation is higher order
  (for the neutral current couplings) or physically unobservable (for the
  charged current ones).}

This factor 2/3 can be
easily understood as follows. At order $v^2/f^2$ we 
can completely describe the SM charged lepton masses
and their couplings to the Higgs boson by means of the following
effective Lagrangian, written in the basis defined by
Eq.~(\ref{SMleptonmasses}), 
\begin{align}
    \mathcal{L}_{\mathit{eff}}=&-\frac{\sqrt{2}}{v}m_{\ell_i}
  \overline{l_{L\,i}} \phi\, \ell_{R\,i}
+\frac{c_{ij}}{f^2} |\phi|^2 \overline{l_{L\,i}} \phi
\,\ell_{R\,j}+\mathrm{h.c.} +\ldots
\\
=&\left[\left(-m_{\ell_i} \delta_{ij} +
  \frac{1}{2\sqrt{2}}\frac{v^3}{f^2}c_{ij}
  \right)
  +\frac{h}{v}\left(-m_{\ell_i} \delta_{ij} +
  \frac{3}{2\sqrt{2}}\frac{v^3}{f^2}c_{ij}
  \right)\right]
\overline{\ell_{L\,i}} \, \ell_{R\,j}  +\mathrm{h.c.} +\ldots~, \nonumber
\end{align}
where $c_{ij}$ are the corresponding (one-loop) Wilson coefficients.
The key point is the relative factor of 3 between the Yukawa coupling and
the mass term at order $v^2/f^2$, originating from the expansion
$(v+h)^3=v^3+3v^2 h + \ldots$ in the dimension 6 operator above.
Due to this factor the mass and Yukawa
matrices are no longer proportional to each other and diagonalizing
the former does not automatically diagonalize the latter. We
can go to the physical basis by means of the usual bi-unitary
transformation
\begin{equation}
  \ell_{L,R\,i}=(U_{L,R})_{ij}\, \ell_{L,R\,j}^{phys}\,,
\end{equation}
where we have emphasized that $\ell_{L,R}^{phys}$ are the
charged leptons in the physical basis and $U_{L,R}$ are
$3\times 3$ unitary matrices that can be written, up to
order $v^2/f^2$, as
\begin{equation}
  U_{L,R}=1+\frac{v^2}{f^2} A_{L,R}\,,
\end{equation}
with $A_{L,R}$ antihermitian matrices. The explicit form of these
matrices can be found in~\cite{delAguila:2000aa} but it is not needed
for the discussion of the off-diagonal terms. Then, the condition that
the mass matrix is diagonal in the physical
basis in particular requires that the coefficients of the off-diagonal
terms of order $v^2/f^2$ cancel in this basis
\begin{equation}
\frac{c_{ij} v}{2\sqrt{2}} + (A_L)_{ij}
  m_{\ell_j} - m_{\ell_i} (A_R)_{ij}=0\,,\quad (i\neq j, \mbox{
  physical basis}), 
\end{equation}
which in turn implies, for the off-diagonal contribution to the Yukawa
coupling,
\begin{align}
  \begin{aligned}
\frac{v^2}{f^2} & \left[  \frac{3\,c_{ij} }{2\sqrt{2}} + (A_L)_{ij}
  \frac{m_{\ell_j}}{v} - \frac{m_{\ell_i}}{v} (A_R)_{ij}\right]
\,h\, \overline{\ell_{L\,i}} \, \ell_{R\,j} \,
+\ldots
\\
&\quad=
 \frac{1}{\sqrt{2}}\frac{v^2}{f^2}c_{ij}
 \,h\, \overline{\ell_{L\,i}} \, \ell_{R\,j} \,
+\ldots,
 \quad (i\neq j, \mbox{ physical basis}).
\end{aligned}\end{align}
Thus we see that the effect of going to the physical
basis just amounts to a simple re-scaling of the
\textit{off-diagonal} Yukawa couplings by a factor $2/3$.

The LFV partial width can therefore be written as
\begin{equation}
\Gamma(h\to \tau^+ \mu^- + \tau^- \mu^+)=\frac{M_h}{16\pi} 
\frac{m_\tau^2 + m_\mu^2}{v^2} \,\frac{4}{9}\,
(|c^{\tau \mu}_L|^2 + |c^{\tau \mu}_R|^2) \ ,
\end{equation}
and its branching ratio 
\begin{equation}
{\rm Br}(h\to \tau \mu) = {\rm Br}(h\to b\bar b)\ 
\frac{\Gamma(h\to \tau^+ \mu^- + \tau^- \mu^+)}{\Gamma(h\to b\bar b)}
\simeq 0.6\ \frac{m_\tau^2}{6 m_b^2}
\,\frac{4}{9}\, (|c^{\tau \mu}_L|^2 + |c^{\tau \mu}_R|^2)\ .
\end{equation}
Now, using Eq. (\ref{AL-AR}), with the mixing matrices 
(the $V$ columns correspond to $e, \mu$ and $\tau$, respectively) 
\begin{equation}
V=\begin{pmatrix}
1 & 0 & 0 \\ 
0 & \cos\theta_1 &  \sin\theta_1 \\
0 & -\sin\theta_1  & \cos\theta_1  
\end{pmatrix}\ , \quad 
W=\begin{pmatrix}
1 & 0 & 0 \\ 
0 & \cos\theta_2 &  \sin\theta_2 \\
0 & -\sin\theta_2  & \cos\theta_2  
\end{pmatrix}\ ,
\end{equation}
we obtain 
\bea
\begin{aligned}
{\rm Br}(h\to \tau \mu) \simeq 
0.2 \times 10^{-6}\ , 
\end{aligned}
\eea 
for $f = 1$ TeV, fixing the heavy gauge boson masses $M_{W_H, Z_H, A_H}$ in Eq. (\ref{MWZ}) 
and $M_{\Phi} \simeq \sqrt 2 M_h f / v$ \cite{Han:2003wu}, 
and $m_{\ell_{H 2,3}} = 1.0, 8.1$ TeV, 
$m_{\tilde{l}_{2,3}} = 10, 50$ TeV, and $\theta_{1,2} = \frac{\pi}{3}, \frac{\pi}{25}$, respectively. 
In general, the LFV Higgs branching ratios tend to be smaller 
when $m_{\tilde{l}_{i}} \sim m_{\ell_{H i}} \sim M_{W_H}$ and there are often large cancellations. 
At any rate, in order to assess the experimentally 
allowed regions in parameter space in the LHT these predictions 
have to be confronted with the corresponding ones with gauge 
bosons, and all of them with the stringent experimental 
limits on LFV processes. We will present such a detailed study 
elsewhere \cite{inpreparation}. 

\section{Summary and conclusions}
\label{Summary}

We have calculated loop induced lepton flavor violating Higgs decays in the Littlest Higgs model with T-parity including all contributions from the T-odd lepton sector. 
We have shown that a finite amplitude is obtained only when all of these contributions are included in contrast to lepton flavor violating processes mediated by gauge bosons where the partners of the right-handed mirror leptons can be decoupled from the spectrum.  
These partners are necessary to cancel the divergence in the Higgs mass introduced by the mirror leptons but are otherwise unnecessary and assumed to be decoupled in previous phenomenological studies.
We have emphasized that these partner leptons can not be decoupled in Higgs decays and furthermore, they introduce a new source of lepton flavor violation via their couplings to the physical pseudo-Goldstone electroweak triplet scalar.

Although this extra source also affects lepton flavor violating processes mediated by gauge bosons, it decouples from these amplitudes in the limit of heavy mass for the partner leptons.
However, if all the partner leptons are kept at the same order as the other T-odd particles, all their contributions are expected to be of similar size. 
This implies that the contributions of partner leptons as well as the pseudo-Goldstone scalar electroweak triplet must also be taken into account when estimating LFV processes mediated by photons and $Z$ bosons. 
Moreover, an assessment about the parameter space of this model allowed by experiment requires the calculation of these new contributions and to perform a global fit to all LFV processes experimentally accessible. 
A detailed discussion of the corresponding limits will be presented elsewhere. 

Barring these further constraints, we find that the $h \rightarrow \tau \mu$ branching ratio can be as large as $\sim 0.2 \times 10^{-6}$ for large mixings and all T-odd particle masses of the order a few TeV.

\acknowledgments

We thank useful discussions and comments by A. David, M. Masip,
R. Pittau, J. Wudka and the anonymous referee.  
This work has been supported in part by the European 
Commission through the contract PITN-GA-2012-316704 (HIGGSTOOLS), by
the Ministry of Economy, Industry and Competitiveness (MINEICO), under grant numbers 
FPA2013-47836-C3-1,2,3-P and FPA2016-78220-C3-1,2,3-P (fondos FEDER), 
and by the Junta de Andaluc{\'\i}a grant FQM 101. P.T. is partially supported by FPA2013-46570.

\appendix
\section{Appendix}
\label{Appendix}

The function $F$ in Eq. (\ref{AL-AR}) can be split depending on the 
fields running in the loop and their common mass dependence in 8 pieces corresponding to the first 8 row groupings in Table \ref{1infinities}, separated 
by horizontal lines:\footnote{
Below, we use $m_{\ell_{H i}}$ everywhere, although in the (charged) diagrams 
exchanging mirror neutrinos, $m_{\nu_{H i}}$ must be understood. (Similarly 
to the case of the function $G$ in Eq. (\ref{G}), where the relevant mass 
is $m_{\tilde{\nu}^c_{j}}$, when applicable.) 
Analogously, we denote by $M_\Phi$ the mass of any scalar triplet component, although they 
can differ by a small amount ${\cal O}(v^2 / f^2)$ after EWSB, which we can 
neglect at the order we work.
} 
\bea
F = \left. F\right|_{W_H} + \left. F\right|_{Z_H} + \left. F\right|_{A_H} + \left. F\right|_{Z_HA_H} 
+ \left. F\right|_{\Phi} + \left. F\right|_{\omega\Phi} + \left. F\right|_{\omega^0\Phi^P} + 
\left. F\right|_{\eta\Phi^P} \ .
\eea
Using the scalar integrals of two and three-point functions in 
Ref. \cite{Passarino:1978jh} (see also Ref. \cite{delAguila:2008zu} for notation; in particular, 
we omit the first three arguments of the three-point functions, which we take equal to $0$ 
for the external fermion momenta and to $Q^2 = M_h^2$ for the Higgs momentum), 
these contributions read  
\bea
\begin{aligned}
\left. F\right|_{W_H} & = 
 - \frac{1}{16} - \frac{1}{16} \frac{m_{\ell_{H i}}^2}{M_{W_H}^2}
  + \frac{1}{2} C_{00}(m_{\ell_{H i}}^2, M_{W_H}^2, M_{W_H}^2)
+ \frac{m_{\ell_{H i}}^2}{24} C_0(m_{\ell_{H i}}^2, M_{W_H}^2, M_{W_H}^2) \\  
& - \frac{1}{2} \frac{m_{\ell_{H i}}^2}{M_{W_H}^2} \left[ 
 \frac{1}{12} B_0(0;M_{W_H}^2, m_{\ell_{H i}}^2) 
  -  C_{00}(M_{W_H}^2, m_{\ell_{H i}}^2, m_{\ell_{H i}}^2) \right]  \\ 
 & + \frac{m_{\ell_{H i}}^2}{2} \left[\left( \frac{1}{3} + \frac{M_{W_H}^2}{m_{\ell_{H i}}^2} \right) 
 C_1(m_{\ell_{H i}}^2, M_{W_H}^2, M_{W_H}^2) \right. \\ 
& \qquad \left. - \left(1+ \frac{1}{2} \frac{m_{\ell_{H i}}^2}{M_{W_H}^2}\right) 
C_1(M_{W_H}^2, m_{\ell_{H i}}^2, m_{\ell_{H i}}^2)
 - \frac{1}{2} C_0(M_{W_H}^2, m_{\ell_{H i}}^2, m_{\ell_{H i}}^2) \right] \ ,
\end{aligned}
\eea
\bea
\begin{aligned}
\left. F\right|_{Z_H} &=
 - \frac{1}{32}
 + \frac{1}{4}C_{00}(m_{\ell_{H i}}^2, M_{W_H}^2, M_{W_H}^2) 
  + \frac{1}{48} m_{\ell_{H i}}^2 C_0(m_{\ell_{H i}}^2, M_{W_H}^2, M_{W_H}^2) \\
&   - \frac{1}{24} \frac{m_{\ell_{H i}}^2}{M_{W_H}^2} \left[\frac{1}{2} 
 B_1(0;m_{\ell_{H i}}^2, M_{W_H}^2) -  B_0(0; M_{W_H}^2, m_{\ell_{H i}}^2) \right] \\
& + ( \frac{1}{8} M_{W_H}^2 + \frac{1}{12} m_{\ell_{H i}}^2 )\ C_1(m_{\ell_{H i}}^2, M_{W_H}^2, M_{W_H}^2) 
\ ,
\end{aligned}
\eea
\bea
\begin{aligned}
\left. F\right|_{A_H} &=
- \frac{1}{4} \frac{M_{A_H}^2}{M_{W_H}^2} \left[ \frac{1}{8}
 -  C_{00}(m_{\ell_{H i}}^2, M_{A_H}^2, M_{A_H}^2)
 - \frac{1}{12} m_{\ell_{H i}}^2 C_0(m_{\ell_{H i}}^2, M_{A_H}^2, M_{A_H}^2) \right] \\
& - \frac{1}{8} \frac{m_{\ell_{H i}}^2}{M_{W_H}^2} \left[
\frac{1}{6} B_1(0;M_{A_H}^2, m_{\ell_{H i}}^2) 
+ B_0(0;M_{A_H}^2, m_{\ell_{H i}}^2) \right] \\
& + \frac{M_{A_H}^2}{M_{W_H}^2} 
\left( \frac{1}{8} M_{A_H}^2 + \frac{1}{12} m_{\ell_{H i}}^2 \right) 
C_1(m_{\ell_{H i}}^2, M_{A_H}^2, M_{A_H}^2) \ ,
\end{aligned}
\eea
\bea
\begin{aligned}
\left. F\right|_{Z_HA_H} &=
\frac{1}{24} \frac{m_{\ell_{H i}}^2}{M_{W_H}^2} \left[ \frac{1}{2} B_1(0;M_{W_H}^2, m_{\ell_{H i}}^2)
 + \frac{1}{2} B_1(0;M_{A_H}^2, m_{\ell_{H i}}^2) + B_0(0;M_{W_H}^2, m_{\ell_{H i}}^2) \right. \\
& \qquad \qquad  +  B_0(0;M_{A_H}^2, m_{\ell_{H i}}^2) \Big] 
 - \frac{1}{4}M_{A_H}^2 C_1(m_{\ell_{H i}}^2, M_{W_H}^2, M_{A_H}^2)  \\
& + \frac{m_{\ell_{H i}}^2}{M_{W_H}^2} 
\left( \frac{M_{W_H}^2}{24}+\frac{M_{A_H}^2}{24} \right) \Big[
    C_1(m_{\ell_{H i}}^2, M_{W_H}^2, M_{A_H}^2)
    + C_0(m_{\ell_{H i}}^2, M_{W_H}^2, M_{A_H}^2) \Big] \ ,
\end{aligned}
\eea
\bea
\begin{aligned}
\left. F\right|_{\Phi} &=
   \frac{1}{16}
 - \frac{1}{2} C_{00}(m_{\ell_{H i}}^2, M_\Phi^2, M_\Phi^2)
 - \frac{1}{24} \frac{m_{\ell_{H i}}^2}{M_{W_H}^2} B_0(0;M_\Phi^2, m_{\ell_{H i}}^2) \ ,
\end{aligned}
\eea
\bea
\begin{aligned}
\left. F\right|_{\omega\Phi} &=
\frac{1}{24} \frac{m_{\ell_{H i}}^2}{M_{W_H}^2}  \left[ B_0(0; M_{W_H}^2, m_{\ell_{H i}}^2)
+ B_0(0;M_\Phi^2, m_{\ell_{H i}}^2) \right.  \\
& \left. + ( M_{W_H}^2 + M_\Phi^2)\ C_0(m_{\ell_{H i}}^2, M_{W_H}^2, M_\Phi^2) \right] \ ,
\end{aligned}
\eea
\bea
\begin{aligned}
\left. F\right|_{\omega^0\Phi^P} &=
\frac{1}{96} \frac{m_{\ell_{H i}}^2} {M_{W_H}^2} \left[ B_0(0;M_{W_H}^2, m_{\ell_{H i}}^2)
 + B_0(0;M_\Phi^2, m_{\ell_{H i}}^2)  \right. \\
&  \left. +  (M_{W_H}^2 + M_\Phi^2)\ C_0(m_{\ell_{H i}}^2, M_{W_H}^2, M_\Phi^2)
 \right] \ , 
\end{aligned}
\eea
\bea
\begin{aligned}
\left. F\right|_{\eta\Phi^P} &=
-\frac{1}{96} \frac{m_{\ell_{H i}}^2} {M_{W_H}^2} \left[ B_0(0;M_{A_H}^2, m_{\ell_{H i}}^2)
 + B_0(0;M_\Phi^2, m_{\ell_{H i}}^2)  \right. \\
&  \left. +  (M_{A_H}^2 + M_\Phi^2)\ C_0(m_{\ell_{H i}}^2, M_{A_H}^2, M_\Phi^2)
 \right] \ .
\end{aligned}
\eea
We neglect terms proportional to $Q^2 = M_h^2$ because they are next order in $v^2/f^2$. 
Corrections proportional to light (SM) lepton masses are also neglected everywhere.

\end{document}